\documentclass[useAMS,articles,fleqn] {mnras}
\usepackage{epsfig}
\usepackage{amsmath}
\usepackage{graphicx}
\usepackage{amssymb}

\title[Variations of the IMF in the Milky Way]{Massive stars reveal variations of the stellar initial mass function in the Milky Way stellar clusters}

\author[Dib et al.]{Sami Dib$^{1,2}$\thanks{E-mail: sami.dib@gmail.com; sdib@nbi.dk}, Stefan Schmeja$^{3,4}$, Sacha Hony$^{5}$\\ 
$^{1}$Niels Bohr Institute \& Centre for Star and Planet Formation, University of Copenhagen, {\O}ster Voldgade 5-7, DK-1350, Copenhagen, Denmark.\\ 
$^{2}$Unidad de Astronom\'{i}a, Departamento de Fisica, Universidad de Atacama, Copayapu 485, Copiapo, Chile.\\
$^{3}$Astronomisches Rechen-Institut, Zentrum f\"{u}r Astronomie der Universit\"{a}t Heidelberg, M\"{o}nchhofstra{\ss}e 12-14, 69120 Heidelberg, Germany.\\ 
$^{4}$Technische Informationsbibliothek, Welfengarten 1b, 30167 Hannover, Germany\\
$^{5}$Institut f\"{u}r Theoretische Astrophysik, Zentrum f\"{u}r Astronomie der Universit\"{a}t Heidelberg, Albert-\"{U}berle-Stra{\ss}e 2, 69120 Heidelberg, Germany.\\
}

\begin{document}
\maketitle

\date{Accepted XXX. Received XXX}

\pagerange{\pageref{firstpage}--\pageref{lastpage}}
\pubyear{2016}
\label{firstpage}

\begin{abstract} 

We investigate whether the stellar initial mass function (IMF) is universal, or whether it varies significantly among young stellar clusters in the Milky Way. We propose a method to uncover the range of variation of the parameters that describe the shape of the IMF for the population of young Galactic clusters. These parameters are the slopes in the low and high stellar mass regimes, $\gamma$ and $\Gamma$, respectively, and the characteristic mass, $M_{ch}$. The method relies exclusively on the high mass content of the clusters, but is able to yield information on the distributions of parameters that describe the IMF over the entire stellar mass range. This is achieved by comparing the fractions of single and lonely massive O stars in a recent catalog of the Milky Way clusters with a library of simulated clusters built with various distribution functions of the IMF parameters. The synthetic clusters are corrected for the effects of the binary population, stellar evolution, sample incompleteness, and ejected O stars. Our findings indicate that broad distributions of the IMF parameters are required in order to reproduce the fractions of single and lonely O stars in Galactic clusters. They also do not lend support to the existence of a cluster mass-maximum stellar mass relation. We propose a probabilistic formulation of the IMF whereby the parameters of the IMF are described by Gaussian distribution functions centered around  $\gamma=0.91$, $\Gamma=1.37$, and $M_{ch}=0.41$ M$_{\odot}$, and with dispersions of $\sigma_{\gamma}=0.25$, $\sigma_{\Gamma}=0.60$, and $\sigma_{M_{ch}}=0.27$ M$_{\odot}$ around these values.

\end{abstract} 


\section{INTRODUCTION}\label{intro}

The initial mass function (IMF) of stars in the Galaxy (i.e., the distribution of the masses of stars at their birth), is of fundamental importance for astrophysics. The IMF controls the efficiency of star formation in molecular clouds (e.g., Zinnecker \& Yorke 2007; Dib et al. 2011a,b; 2013), the size distribution of protoplanetary disks in stellar clusters (e.g., Vincke et al. 2015), the radiative and mechanical feedback from stars into the interstellar medium (e.g., Dib et al. 2006; Martizzi et al. 2016) and the dynamical and chemical evolution of galaxies (e.g., Boissier \& Prantzos 1999). In the Milky Way, as in other galaxies, stars form mostly, if not exclusively, in clusters and associations (e.g., Carpenter 2000, Lada \& Lada 2003, Hony et al. 2015). As clusters age, the expulsion of gas by stellar feedback as well as dynamical interactions between stars and binary systems in the cluster soften its gravitational potential, leading to its expansion and to its partial or total dissolution into the field of the galaxy (e.g., Goodwin \& Bastian 2006; Pfalzner \& Kaczmarek 2013). The mass function of stars in the field of a galaxy is thus the convolution of the galaxy$'$s cluster formation history with the stars from dissolved clusters and the stars that have been ejected from surviving clusters. In our Galaxy, the present day stellar mass function, uncorrected for the binary population, rises from the brown dwarf and low stellar mass regime until it peaks at $\approx 0.3-0.5$ M$_{\odot}$ after which it declines steeply in the intermediate-to-high mass regime (e.g., Miller \& Scalo 1979; Scalo 1986; Kroupa 1993; Chabrier 2003; Bochanski et al. 2010; Rybizki \& Just 2015). Several distribution functions are used to describe its shape, such as a multi-component power-law (Kroupa 2001), a lognormal coupled to a power-law beyond 1 M$_{\odot}$ (Chabrier 2005), a tapered power law (de Marchi et al. 2010; Parravano et al. 2011), an order-3 Logistic function (Maschberger 2013), or a modified lognormal (Basu et al. 2015). 

An outstanding question is whether there are significant variations in the shape of the IMF among stellar clusters in the Milky Way and how well the IMF of each cluster resembles the mass function of stars in the Galaxy (e.g., Elmegreen 2004; Scalo 2005; Dib 2014a). Stars in young clusters have roughly the same age, metallicity, and are located at the same distance. Thus, one can presume that their observed present day mass functions (PDMFs\footnote{In the remainder of the paper, we will refer to the PDMF, especially to that of young clusters as being the IMF. However, it should always be kept in mind that we are dealing here with PDMFs.}) are a fair representation of their IMFs. Probing the universality of the IMF among stellar clusters in the Milky Way and in other galaxies is one of the most challenging issues in modern astrophysics. For a Galactic star formation rate (SFR) between $0.5$ and $1.5$ M$_{\odot}$ yr$^{-1}$, the Galaxy is expected to form a few tens to a fews hundred thousands of clusters in a period of $\sim 10-12$ Myrs\footnote{The exact numbers depend primarily on the SFR, the exponent of the initial cluster mass function (ICLMF), an the lower and upper mass cutoffs of the ICLMF. See \S.~\ref{appendixa} for more quantitative estimates}. The IMF has been derived in the Galaxy and in the Magellanic Clouds by many groups for a small fraction of this total number and usually for individual young clusters and associations as well as for more evolved open clusters (e.g., Massey et al. 1998; Preibisch et al. 2002; Liu et al. 2003; Luhman 2004,2007; Moraux et al. 2004; Selman \& Melnick 2005; Bouvier et al. 2008; Liu et al, 2009; Sung \& Bessel 2010; Ojha et al. 2010; Delgado et al. 2011; Gennaro et al. 2011; Lodieu et al. 2011; Alves de Oliveira et al. 2012, Mallick et al. 2014; Maia et al. 2016 among many others). The comparison of the parameters that describe the shape of the IMF between these works is not straight forward. Observations of stellar clusters have been carried out using different telescopes with different sensitivities, and different methods are employed to reduce the data and to correct for the effects of extinction and stellar incompleteness. The conversion of measured stellar fluxes into masses is also performed using different stellar evolutionary tracks (see interesting discussions in Scalo 1998 and Massey 2011 on this topic). Based on the comparison of a relatively small number of clusters compiled from these observations, there are claims that within the uncertainties, the shape of the IMF of some clusters are similar, at least in the intermediate to high-mass stellar regime (e.g., Bastian et al. 2010; Offner et al. 2014). However, there are also a few other studies in which the parameters of the IMF have been derived using a more homogeneous approach and that show significant cluster-to-cluster variations (e.g., Sharma et al. 2008; Massey 2011; Scholz et al. 2013; Dib et al. 2014a; Lim et al. 2015; Weisz et al. 2015). In principle, a direct assessment of the universality of the IMF could be achieved by constructing the IMF for a large number of Galactic and extragalactic young clusters across the entire stellar mass range. This is however beyond the reach of current observational programs. Surveys that contain a large number of clusters such as the PHAT survey of the Andromeda galaxy (85 clusters) are sensitive only to the intermediate-to-high mass stellar content of the clusters (stars with masses M$_{*} \gtrsim  2$ M$_{\odot}$) and thus can only make statements about the IMF in this mass regime (Weisz et al. 2015). As in the case of Galactic clusters (e.g., Sharma et al. 2008; Dib 2014a), the findings of Weisz et al. (2015) indicate values of the slope of the IMF in the intermediate-to-high mass regime that do not overlap within the $1\sigma$ confidence intervals (see Figure 4 in their paper), and that are, for many of them, not compatible with the values of the parameters for the Galactic field stellar mass function. 

In this paper, we propose an alternative method to uncover the range of variation of the parameters that describe the IMF for the populations of young clusters ($ \lesssim 12$ Myrs) in the Milky Way. The method is based on the fact that the number statistics of massive stars in Galactic clusters is very sensitive to the underlying distribution of the IMF parameters in the clusters. The method relies exclusively on the high mass content of the clusters, but is able to yield information on the distributions of parameters of the IMF over the entire stellar mass range. This is achieved by appropriately comparing the fractions of single and lonely O stars in a recent catalog of the Milky Way clusters (the MWSC catalog; Kharchenko et al. 2013; Schmeja et al. 2014) with a large library of simulated clusters built with various distribution functions of the IMF parameters. The simulated synthetic clusters include corrections for the binary population, stellar evolution and sample incompleteness. In \S.~\ref{method}, we discuss the essential aspects of the method that is employed to compare models and observations and in \S.~\ref{observations}, we briefly present the observational data. The models of synthetic clusters are presented in \S.~\ref{models}, and the comparison to the observation is performed in \S.~\ref{comparison}. In \S.~\ref{otherwork} we compare our approach to previous work on closely related topics and in \S.~\ref{discussion} we discuss our results in connection to the physical processes that may lead to variations of the IMF. Finally, we present our conclusions in \S.~\ref{conclusions}. 

\section{Method}\label{method}

The number statistics of massive stars in clusters is very sensitive to the underlying distribution of the IMF parameters in the clusters. In this work, we want to make use of the massive stellar population in young Galactic stellar clusters in order to infer the distribution of the parameters that describe the shape of the IMF across the entire stellar mass range. This can be achieved by comparing the fractions of single and lonely massive O stars in a recent catalog of the Milky Way clusters (the MWSC catalog; Kharchenko et al. 2013; Schmeja et al. 2014) with a large library of simulated clusters built with various distribution functions of the IMF parameters. Since in the simulated clusters, we are populating their system IMFs, a star as defined in this work could be an individual star or a binary system. Thus, an O star in the cluster could be a single star or one/both components of a binary system with a mass that is $\geq 15$ M$_{\odot}$. An O star in a cluster is  called "single" if it is the only living star, or a binary system with any of its components, that has a mass $\geq 15$ M$_{\odot}$ in the cluster. The fraction of single O stars\footnote{For simplicity, we will use the term "star" to define both individual stars or binary systems} in a population of clusters is thus given by:

\begin{equation}
f_{O,single}=\frac{N_{O,single}}{N_{O}},
\label{eq1}
\end{equation}

\noindent where $N_{O,single}$ is the total number of single O stars and $N_{O}$ is the total number of O stars in all clusters. We also measure the fraction of "lonely" O stars in the clusters. A lonely O star in a cluster is a single O star with the additional constraint that the next massive system in the cluster is less massive than $10$ M$_{\odot}$ (i.e., absence of high mass B stars with masses between 10 and 15 M$_{\odot}$). The fraction of lonely O stars is given by:

\begin{equation}
f_{O,lonely}=\frac{N_{O,lonely}}{N_{O}},
\label{eq2}
\end{equation}

\noindent where $N_{O,lonely}$ is the total number of lonely stars in the clusters. The basic idea of the method relies on the comparison of $f_{O,single}$ and $f_{O,lonely}$ measured for a recent catalogue of stellar clusters in the Milky Way with those derived for populations of stellar clusters that are generated with various prior functions for the distributions of the IMF parameters. A number of important corrections have to be applied to the zero-age synthetic clusters and to their stellar content before they can be compared to the observations. The description of the observational sample of clusters is given in \S.~\ref{observations}, while the synthetic models of clusters are described in \S.~\ref{models}. 

\section{Observational catalog of clusters}\label{observations}

The observations used in this paper come from the Milky Way Stellar Clusters survey (MWSC) (Kharchenko et al. 2012; Kharchenko et al. 2013; Schmeja et al. 2014)\footnote{The full list of cluster parameters is available at http://vizier.cfa.harvard.edu/viz-bin/VizieR?-source=J/A+A/558/A53}, which lists $\sim 3200$ clusters with ages between $\sim 1$ Myr and $\sim 7$ Gyr. The clusters are detected as density and velocity enhancements in the Two Micron All Sky Survey (2MASS) (Skrutskie et al. 2006) and the proper motions PPMXL survey (R\"{o}ser et al. 2010). The survey includes clusters up to a distance of $\sim 10$ kpc from the position of the Sun with a substantial fraction of the clusters being located within a distance $\lesssim 1.8$ kpc (see Figure 1 in Schmeja et al. 2014). The files in the catalog list the B, V, and J, H, K magnitudes of each star in each cluster present in the catalog, along with other properties such as position, age, cluster membership probability, proper motions, and, when available, the spectral type. The resolved stellar content of each cluster are however not corrected for the effects of the binary population. In this work, we are interested in clusters that could, based on their age, harbor high-mass stars ($M_{*} \geq 15$ M$_{\odot}$), which implies clusters younger than $\tau_{15}\approx 12.3$ Myrs, where $\tau_{15}$ is the duration of the Hydrogen and Helium burning phases for a star with a mass of 15 M$_{\odot}$ (Ekstr\"{o}m et al. 2012). This brings down the number of clusters useful for our purposes in the MWSC to $N_{\rm MWSC,cl}=341$. The individual stellar masses are estimated from the relation between ${\rm log} M_{*}$ and the absolute visual magnitude $M_{V}$ given by Schilbach et al. (2006). The absolute magnitude $M_{V}$ is computed from the apparent magnitude $m_{V}$, the distance, and the extinction $E_{B-V}$ which are all listed in the MWSC (Kharchenko et al 2013). Out of the total 341 clusters, 175 of them contain at least one O star ($M_{*} \ge 15$ M$_{\odot}$). The number of single and lonely O stars in the observational sample is $N_{O,single}=89$ and $N_{O,lonely}=29$, respectively, and the total number of O stars is $N_{O}=688$. Thus, the fractions of single and lonely O star in the MWSC catalog measured using Eq.~\ref{eq1} and Eq.~\ref{eq2} are $f_{O,single}({\rm MWSC})=12.9 \%$ and $f_{O,lonely}({\rm MWSC})=4.2\%$.  

\section{Models}\label{models}

\subsection{Generating populations of zero-age clusters}

The possibility of detecting massive O stars in the Galaxy is bound by the relatively short lifetime of these stars and by the value of the Galactic star formation rate (SFR). The models of stellar clusters that are compared to the observations are generated in the following way: assuming that all stars form in clusters, the total mass contained in the young  population of Galactic stellar clusters that are likely, based on their age, to contain O stars ($M_{*} \geq 15$ M$_{\odot}$) is given by: 

\begin{equation}
\Sigma_{cl}=\int_{0}^{\tau_{15}} {\rm SFR}(t)\times dt, 
\label{eq3}
\end{equation}

\noindent where ${\rm SFR}(t)$ is the time dependent star formation rate over the last $\tau_{15}$ timescale of the lifetime of the Galaxy. The Galactic SFR over such a relatively short period of time can be assumed to be constant\footnote{The instantaneous Galactic SFR cannot be in reality a constant (i.e., the birth of single massive star in the Galaxy will boost the SFR), but the assumption is made here that the time dependent Galactic SFR will fluctuate around the values chosen in this work (see also da Silva et al. 2012 for further discussion on this issue).} and $\Sigma_{cl}$ can be approximated by $\Sigma_{cl} \approx {\rm SFR} \times \tau_{15}$. We consider three values of the Galactic SFR of 0.68, 1, and 1.45 M$_{\odot}$ yr$^{-1}$ that are the lower, central, and upper estimates obtained from the count of young stellar objects in the GLIMPSE survey of the Galactic plane (Robitaille \& Whitney 2010). The individual cluster masses are stochastically sampled from the mass reservoir $\Sigma_{cl}$ using an initial cluster mass function (i.e., the mass function of clusters at their birth, ICLMF). The ICLMF is taken to be a power-law, between the minimum and maximum cluster masses of $M_{cl,min}$ and $M_{cl,max}$, and is given by:

\begin{equation}
\frac{dN_{cl}}{dM_{cl}}=A_{cl} \times M_{cl}^{-\beta},
\label{eq4}
\end{equation}

\noindent where $A_{cl}$ is the normalization constant given by:

\begin{equation}
\int_{M_{cl,min}}^{M_{cl,max}}A_{cl}\times  M_{cl}^{-\beta+1} dM_{cl}=\Sigma_{cl}.
\label{eq5}
\end{equation}

We fix $M_{cl,max}$ at $5\times10^{4}$ M$_{\odot}$ which is the mass of the most massive clusters in the Milky Way (e.g., Figer et al. 1999; Dib et al. 2007; Ascenco et al. 2007; Harayama et al. 2008; Clark et al. 2009) and explore values of  $M_{cl,min}=50$ (fiducial), $20$, and $10$ M$_{\odot}$. Our fiducial value of $\beta$ is 2 as this is in agreement with the slope of the cluster mass function at intermediate- to high cluster masses in nearby galaxies (e.g., Elmegreen \& Efremov 1997; Zhang \& Fall 1999; Hunter et al. 2005; de Grijs \& Anders 2006; Selman \& Melnick 2008; Larsen 2009; Chandar et al. 2010; Fall \& Chandar 2012) and with theoretical expectations (Elmegreen 2006; Dib et al. 2011; Dib 2011a,b; Dib et al. 2013). We also consider cases with $\beta=2.2$ and $\beta=1.8$. For each cluster with an assigned mass $M_{cl}$, the masses of star-systems (i.e., individual stars or binary systems) in the clusters are randomly sampled using a tapered power law (TPL) distribution function (de Marchi et al. 2010; Parravano et al. 2011). Without any assigned binary fraction, a stellar "system" of mass $M_{*}$ can correspond to an individual star or to a binary system. The TPL function is given by

\begin{equation}
\frac{dN_{*}} {d{\rm log} M_{*}}=A_{*}\times M_{*}^{-\Gamma}\left\{1-\exp\left[-\left(\frac{M_{*}}{M_{ch}}\right)^{\gamma+\Gamma}\right] \right\},
\label{eq6}
\end{equation} 

\noindent where $dN_{*}$ is the number of stellar systems with the logarithm of their masses between ${\rm log}M_{*}$ and ${\rm log} M_{*}+ d{\rm log}M_{*}$, and $A_{*}$ is the normalization coefficient which is given by:
 
 \begin{equation}
 \int_{M_{*,min}}^{M_{*,max}}A_{*}\times M_{*}^{-\Gamma}\left\{1-\exp\left[-\left(\frac{M_{*}}{M_{ch}}\right)^{\gamma+\Gamma}\right] \right\}
dM_{*}=M_{cl}.
 \label{eq7}
 \end{equation}
 
\begin{figure}
\begin{center}
\epsfig{figure=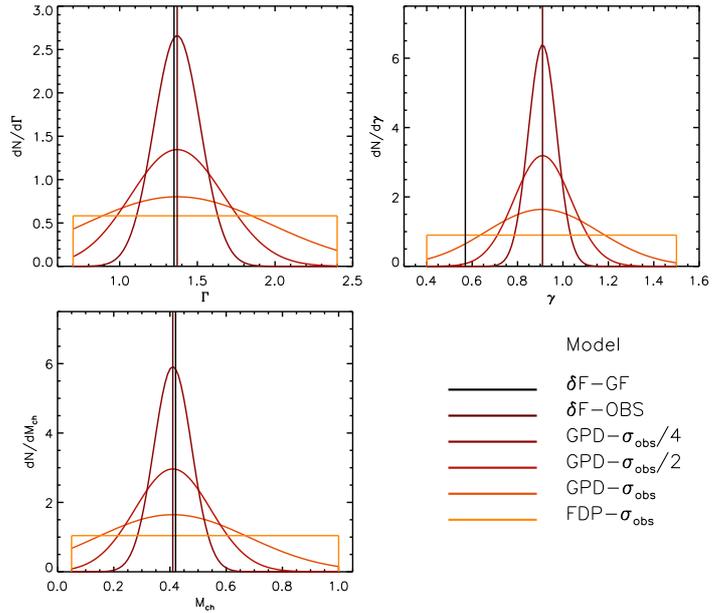,width=\columnwidth}
\end{center}
\vspace{0.3cm}
\caption{The figure displays the probability distribution functions of the three parameters that describe the IMF used in this work. The acronym $\delta$F-GF refers to delta functions of the parameters at the positions of the Galactic field values, whereas $\delta$F-OBS refers to delta functions located at the mean values of the parameters derived by Dib (2014a) for a sample of 8 young Galactic stellar clusters. The cases GPD-$\sigma_{obs}$, GPD-$\sigma_{obs}/2$, and GPD-$\sigma_{obs}/4$ correspond to cases with a Gaussian probability distribution of the IMF parameters whose half-width is related to $1$, $0.5$, and $0.25$ the values of the dispersion of each parameter in the sample of Dib (2014a). FDP-$\sigma_{obs}$ corresponds to a case where the probability distribution function of each parameter is given by a boxcar function whose width is given by $2 \times\sigma_{obs}$. The lower and upper truncations for each of the parameters correspond to the lower and upper limits of these parameters derived by Dib (2014a).}
\label{fig1}
\end{figure}
 
The TPL function describes the IMF with only three parameters, the slope in the low mass regime $\left(\gamma \right)$, the slope in the intermediate-to-high mass regime $\left(\Gamma\right)$, and the characteristic mass $\left(M_{ch}\right)$. The minimum stellar mass, $M_{*,min}$, is always taken to be $0.02$ M$_{\odot}$. The maximum stellar mass, $M_{*,max}$, is either given by ${\rm min}[M_{cl}, 150$ M$_{\odot}$] (corresponding to the case of stochastic sampling) or is dictated by a cluster mass-maximum stellar mass relation ($M_{cl}-M_{*,max}$) relation proposed by Vanbeveren (1982) and later by Weidner \& Kroupa (2004). We test models in which the distributions of the parameters ($\Gamma$, $M_{ch}$, $\gamma$) are either given by delta function (all clusters have the same value of the parameters), Gaussian functions, or boxcar functions. A recent study found that the mean values of the parameters among a relatively small sample of young Galactic stellar clusters are $\Gamma_{obs}=1.37, \gamma_{obs}=0.91$ and $M_{ch,obs}=0.41$ M$_{\odot}$ with standard deviations of  $\sigma_{\Gamma_{obs}}=0.60$, $\sigma_{\gamma_{obs}}=0.25$, and $\sigma_{M_{ch,obs}}=0.27$ M$_{\odot}$, respectively (Dib 2014a). When drawing the parameters from Gaussian distributions, the distributions are always centered on these observed mean values. We test dispersions of the Gaussian distributions of  $\left(\sigma_{\Gamma_{obs}},\sigma_{\gamma_{obs}},\sigma_{M_{ch}} \right)$, $\left(\sigma_{\Gamma_{obs}},\sigma_{\gamma_{obs}},\sigma_{M_{ch}} \right)/2$, and $\left(\sigma_{\Gamma_{obs}},\sigma_{\gamma_{obs}},\sigma_{M_{ch}} \right)/4$, and apply lower and upper cutoffs of $\left(0.4,1.5\right)$ for $\gamma$, $\left(0.70,2.4\right)$ for $\Gamma$, and $\left(0.05,1\right)$ M$_{\odot}$ for M$_{ch}$, which correspond to the lower and upper limits derived by Dib (2014a). These models are labeled GPD-$\sigma/i$ (for Gaussian Probability Distributions, where i=1, 2, or 4). They are contrasted with other models in which the distributions of the IMF parameters are delta functions that are either located at the values of the parameters derived by Dib (2014a), or the Galactic field values that are given by $\Gamma_{field}=1.35$, $\gamma_{field}=0.57$, and $M_{ch,field}=0.42$ M$_{\odot}$ (Parravano et al. 2011). These families of models are labeled $\delta$F-OBS and $\delta$F-GF (Delta Function Observations and- Galactic Field, respectively). We also test flat probability distributions (FPD). These are described by boxcar functions between $\left(0.4,1.5\right)$ for $\gamma$, $\left(0.7,2.4\right)$ for $\Gamma$, and $\left(0.05,1\right)$ M$_{\odot}$ for $M_{ch}$. The distribution functions of the parameters of the IMF for all of these models are displayed in Fig.~\ref{fig1}.

\begin{figure}
\begin{center}
\epsfig{figure=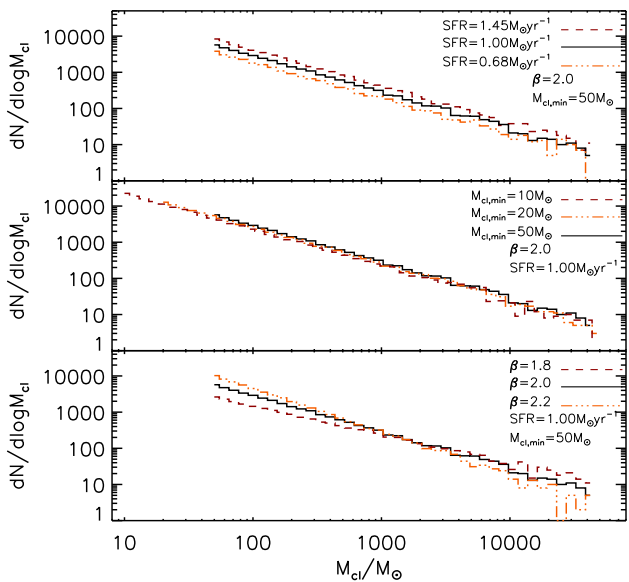,width=0.8\columnwidth}
\end{center}
\vspace{0.5cm}
\caption{The initial cluster mass function (ICLMF) for various values of its parameters. Several realizations of the ICLMF with various values of the star formation rate (top panel), the minimum cluster mass $\left(M_{cl,min}\right)$, and the exponent of the power-law function that describes the ICLMF ($\beta$). The logarithmic bin size is ${\rm log}(M_{cl}/{\rm M}_{\odot})=0.075$.}
\label{fig2}
\end{figure}

\begin{figure}
\begin{center}
\epsfig{figure=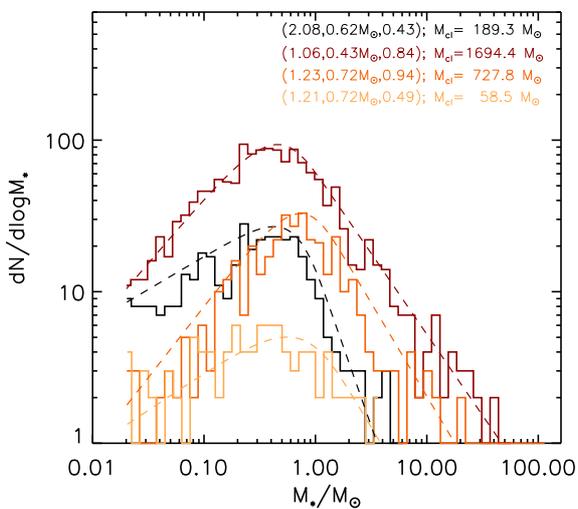,width=\columnwidth}
\end{center}
\caption{Realizations of the initial stellar mass function (IMF), with different permutations of its parameters ($\Gamma, M_{ch} , \gamma)$. The figure displays the shape of the IMF for four permutations of its parameters that are drawn from broad distributions (here from one of the realizations with the GPD-$\sigma_{obs}$ parameter distributions). The inset displays the set of parameters $(\Gamma, M_{ch}, \gamma)$ of the selected clusters, followed by the clusterÕs mass, $M_{cl}$. The logarithmic bin size is ${\rm log}(M_{cl}/{\rm M_{\odot}})=0.075$. The dashed lines are over-plots to the generated data of the continuous form of the IMF generated with the corresponding set of parameters.}
\label{fig3}
\end{figure}

Fig.~\ref{fig2} displays a few examples of the generated ICLMFs with various permutations of the Galactic SFR, the exponent of the ICLMF ($\beta$), and the lower mass cutoff in cluster masses ($M_{cl,min}$), and Fig.~\ref{fig3} displays the system IMFs for a few selected clusters drawn from one of the realization of the ICLMF in the GPD-$\sigma_{obs}$ family of models (i.e., with varying IMFs). Additional technical details on the sampling of the ICLMF and of the IMF are presented in App. \S.~\ref{appendixa}.

\subsection{Assigning ages to simulated clusters}\label{ages}

\begin{figure}
\begin{center}
\epsfig{figure=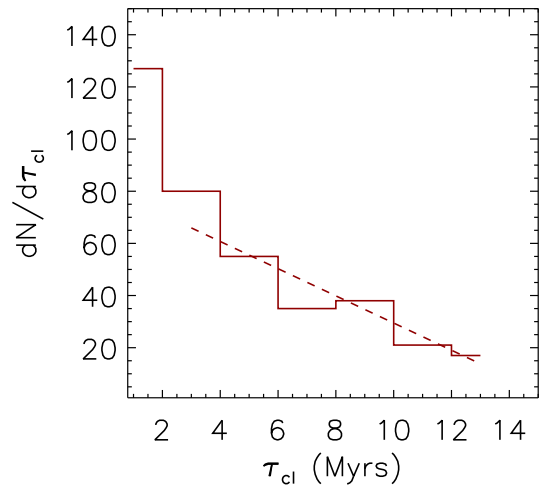,width=\columnwidth}
\end{center}
\caption{The age distribution of young clusters in the MWSC catalog. The bins in clusters ages, $\tau_{cl}$, are linear with a bin size of $\Delta (\tau_{cl})=2$ Myrs. . Cluster ages have been derived using the isochrones based of the Padova stellar evolutionary tracks (Girardi et al. 2002). A linear fit to the cluster age distribution for clusters with ages $\ge 2$ Myrs is shown with the dashed purple line.}
\label{fig4}
\end{figure}

The synthetic clusters that are generated for each realization of the ICLMF are initially zero-age clusters. Before the stellar populations of the clusters can be corrected for the binary fraction and stellar evolution and then compared with the clusters of the MWSC survey, the clusters have to be assigned ages that are compatible with the observed age distribution of the clusters in the MWSC catalog. Fig.~\ref{fig4} displays the age distribution of the young clusters (with ages $\tau_{cl} < 15$ Myrs) in the MWSC catalog. These ages were computed by Kharchenko et al. (2012) using the Padova web-server CMD2.2\footnote{http://stev.oapd.inaf.it/cgi-bin/cmd }, based on the Marigo et al. (2008) calculations for an adopted metallicity of Z = 0.019. The figure shows that there is a mild decline in the number of clusters as a function of cluster age. This is most probably due to the effect of cluster disruption following the onset of gas expulsion which occurs on timescales of a few $10^{5}$ to a few $10^{6}$ yr depending on the clusters masses and star formation history (e.g., Dib et al. 2013). We fit the distribution of cluster ages with a linear function given by (i.e., dashed purple line in Fig.~\ref{fig4}):

\begin{equation}
F(\tau_{cl})=\frac{dN}{d\tau_{cl}}=(-5.21\pm0.72) \times \tau_{cl}+(81.5\pm7.3). 
\label{eq8}
\end{equation}

The bin centered at 1 Myr (i.e., clusters with ages $< 2$ Myrs) is excluded from the fit, as it is particularly difficult to assign accurate ages to very young embedded clusters. The function $F(\tau_{cl})$ can be normalized by requiring that $B_{cl} \int_{0}^{\tau_{15}} F(\tau_{cl}) d\tau_{cl}$=1, where $B_{cl}$ is the normalization constant. The normalized form of Eq.~\ref{eq8} (i.e., $B_{cl}\times F(\tau_{cl})$) is used as a probability distribution function from which the ages of the synthetic clusters in our models are drawn between 0 and $\tau_{15}=12.3$ Myrs. It is important to note that the exact shape of the age distribution has almost no influence on our result. This is because there are no imposed age-mass relation, and thus all cluster masses are well represented at all ages. 

\subsection{Correcting for the effects of binary population and stellar evolution}\label{binary}

In oder to properly count the numbers of O stars in the clusters (whether single, lonely, or neither), we have to account for the effect of stellar evolution. O stars whose Hydrogen+Helium burning phases are shorter than the assigned age of their parent cluster would have turned into stellar remnants (i.e., stellar black holes) and are thus removed from the statistics. The correction for the effects of stellar evolution must be preceded by a correction due to the binary population in the clusters. For each star (i.e., a star-system) with a mass $M_{*} \geq 2$ M$_{\odot}$, we assign a binarity probability that is based on the observed binary fraction measured for a large number of systems in the Galaxy (Chini et al. 2012). For star-systems with masses $\geq 15$ M$_{\odot}$, we assign a binary probability of $P_{bin}=0.82$ which is the mean of the binary fractions for stars with $\geq 15$ M$_{\odot}$, whereas for star-systems in the mass range $2 {\rm M}_{\odot}\leq M_{*} \leq 15{\rm M}_{\odot}$, the binary probability decreases linearly with decreasing mass (Chini et al. 2012). The fit to the observational data is this mass regime is given by $P_{bin}=0.047M_{*}+0.052$. 

The mass ratios of the secondary to the primary stars ($q=M_{2}/M_{1}$) in massive binary systems ($\geq 15$M$_{\odot}$) are randomly drawn from a flat probability distribution following most up to date observational evidence in massive star forming regions such as the Cygnus OB2 associations (Kobulnicky et al. 2014). For binary systems in the B-type stars mass range ($ 2 \leq M_{*}/{\rm M_{\odot}} \leq 15$), the mass ratios are drawn from a mass-ratio distribution that is slightly peaked towards low $q$ values in agreement with the observational measurements in the Sco OB2 association (Shatsky \& Tokovinin 2002). For each of the primary and secondary stars that fulfill $M_{1} \geq 15 {\rm M}_{\odot}$ or $M_{2} \geq 15 {\rm M}_{\odot}$, their Hydrogen+Helium burning lifetime is compared to the age of its cluster ($\tau_{cl}$ ). If the primary star is ÒaliveÓ ($\tau_{M1} \geq \tau_{cl}$), the system is included in the statistics with the system mass $M_{*}$  being substituted by $M_{1}$. Whenever $\tau_{M_{1}} < \tau_{cl}$, the star is considered to have exploded as a supernova. If the secondary star is an O star and it is still alive ($\tau_{M2} \geq \tau_{cl}$), $M_{2}$ is used as the system mass. If both stars have already exploded as supernovae or if both are less massive than $15$ M$_{\odot}$, the system is removed from the statistics. It is important to point out that time dependent effects such as accretion processes by Roche-lobe overflow in massive binaries are not taken into account in this work. Such an effect would depend on the orbital separation of the components of the binary system. Kobulnicky et al. (2014) found that $\approx 80\%$ of the O stars in Cyg OB2 have separations smaller than 1 AU. Therefore they are potential candidates of mass interchange processes during their evolution. If such a period distribution also applies to lower masses stars, some of the low mass stars may accrete enough mass from their massive primaries and become rejuvenated O stars (e.g., Vanbeveren 2009). This would lead to an enhancement of the O star population. However, the potential generation of additional O stars in close binaries does not necessarily imply an enhancement of the fraction of single (or lonely) O stars. This is because in our approach we count binary systems in which both stars are living O stars (for example one by birth, and one by accretion in the binary) as one when it comes to counting  the numbers of single and lonely O, and the total number of O stars. 

\begin{figure*}

\begin{center}
\epsfig{figure=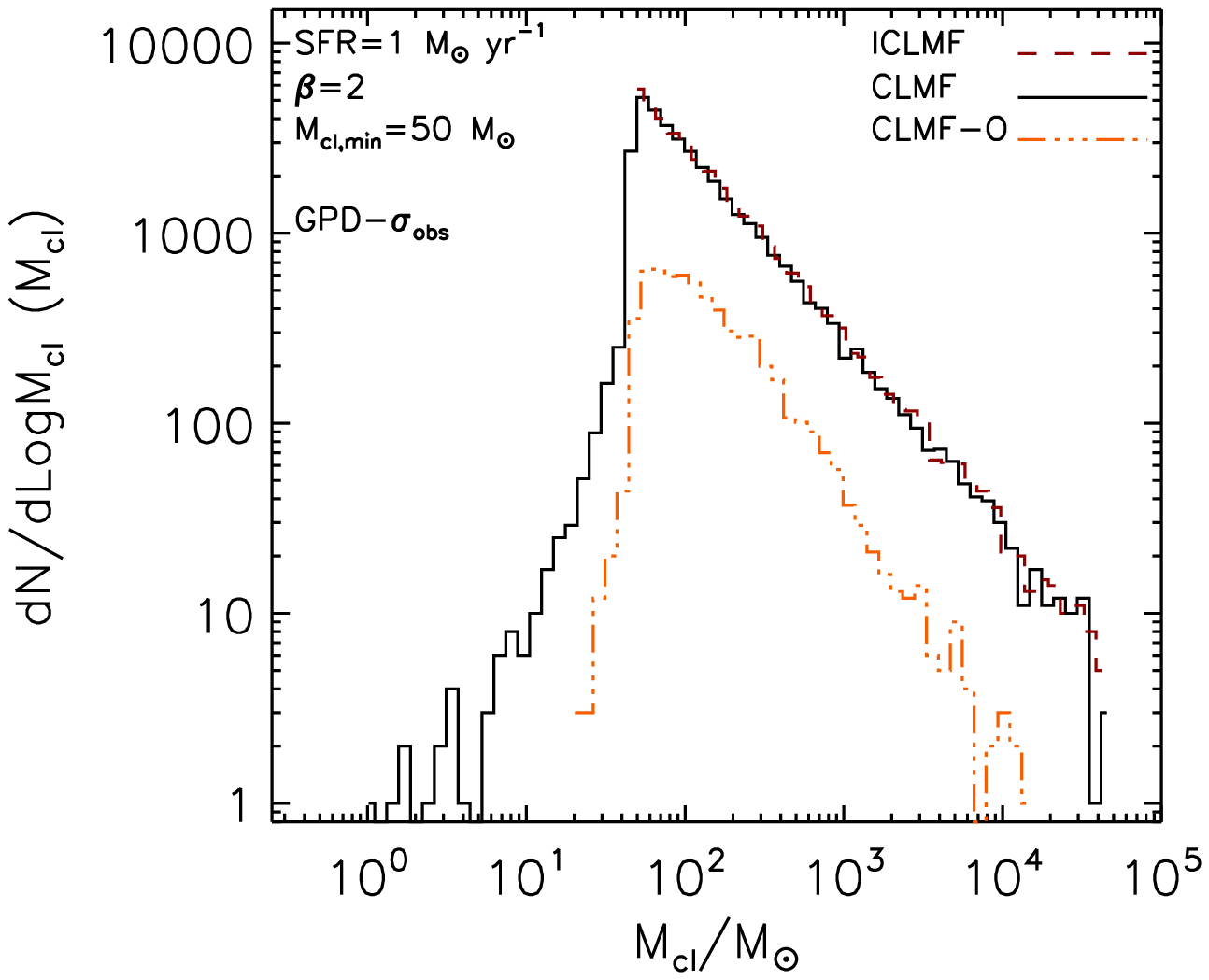,width=0.475\textwidth}
\epsfig{figure=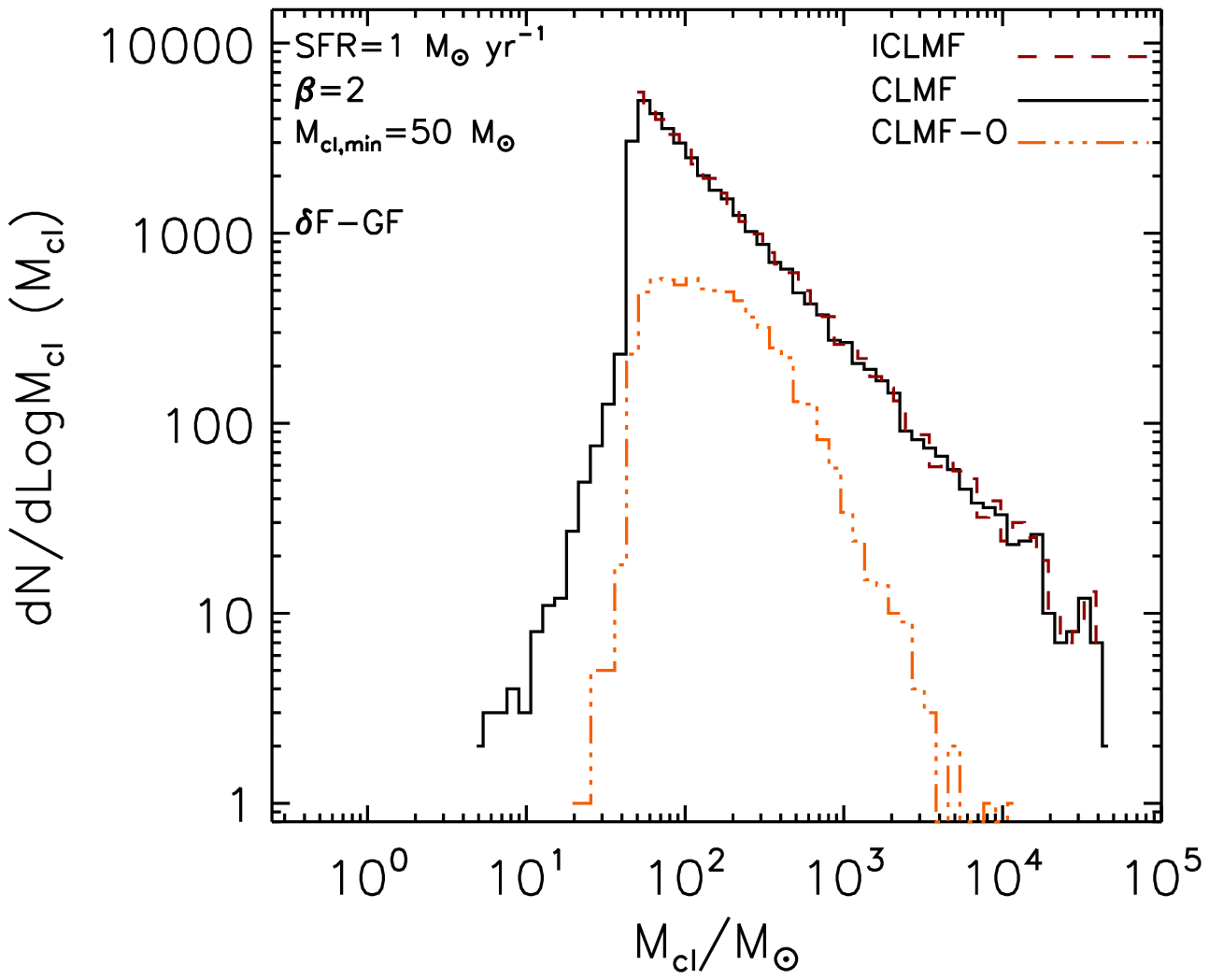,width=0.475\textwidth}
\end{center}
\caption{The initial, present-day, and single O-star cluster mass functions in two realizations of the ICLMF. The left panel displays a case in which the set of three parameters that describe the IMF of each cluster are each randomly drawn from a GPD-$\sigma_{obs}$ probability distribution function whereas the right panel displays a case in which the set of three parameters that describe the IMF is similar to the values of the parameters for the Galactic field mass function. All other parameters are set to $M_{cl,min}=50$ M$_{\odot}$, $M_{*,min}=0.02$ M$_{\odot}$, $M_{*,max}=150$ M$_{\odot}$, and $\beta=2$. The figure displays the initial cluster mass function (ICLMF) and the present-day cluster mass function (CLMF). The CLMF is the conversion of the ICLMF after each individual cluster has been assigned an age, and has been corrected for the effects of the stellar binary fraction and stellar evolution. The number distribution of clusters that contain single O stars (with $M_{*} \geq 15$ M$_{\odot}$ (CLMF-O), peaks at a few tens of stellar masses and most of the clusters in the CLMF-O have masses $ \lesssim 400$ M$_{\odot}$.}
\label{fig5}
\end{figure*}

 Fig.~\ref{fig5} displays two examples of the ICLMF (magenta line), of the corresponding present day cluster mass functions (CLMF) after correcting for binarity and stellar evolution (black line), and of the mass functions of clusters that contain single O-star systems (CLMF-O, triple dot-dash orange line). The left panel corresponds to a case in which the set of three parameters that describe the IMF of each cluster are each randomly drawn from a GPD-$\sigma_{obs}$ probability distribution function whereas the right panel displays a case in which the set of three parameters that describe the IMF is similar to the values of the parameters for the Galactic field mass function (i.e., $\delta$F-GF). The two examples displayed in Fig.~\ref{fig5} are representative of the CLMF and CLMF-O that are obtained for any realization with the same family of IMF models. In both realizations shown here, the other parameters are set to $\beta=2$, SFR=1 M$_{\odot}$ yr$^{-1}$, and $M_{cl,min}=50$ M$_{\odot}$. In both cases, almost all single O stars reside in clusters whose masses are $\leq 400-500$ M$_{\odot}$. A noticeable difference between the two cases in Fig.~\ref{fig5} is that in the case where the set of the three parameters of the IMF ($\Gamma$, $M_{ch}$,$\gamma$) are randomly drawn for each cluster from GPD-$\sigma_{obs}$ distribution functions, there is a slower decrease in the fraction of the clusters that harbor single O stars at high cluster masses (i.e., ratio of CLMF-O to CLMF) in contrast to the case where the three parameters of the IMF assigned to the clusters are identical (i.e., $\delta$F-GF). This is due to the fact that when the IMF parameters are sampled from broad distribution functions, a significant fraction of the clusters will be assigned steep slopes in the intermediate- to high stellar mass range. Massive clusters with a steep slope in the intermediate- to high stellar mass range are more likely to harbor single O stars. This is in contrast to the case with an identical Galactic field-like IMF assigned to all clusters (i.e., $\delta$F-GF) and where the fraction of clusters that harbor single O stars drops quickly as a function of cluster mass, at high cluster masses (Fig.~\ref{fig5}, right panel). Models that have distributions of the IMF parameters that are intermediate between the two models displayed in Fig.~\ref{fig5} (i.e., such as models GPD-$\sigma_{obs}$/2 and GPD-$\sigma_{obs}/4$.) lead to CLMF-O distributions that have intermediate slopes in the high mass regime between those of the $\delta$F-(GF or OBS) models and the GPD-$\sigma_{obs}$ models.
 
\subsection{Correcting for the effects of cluster incompleteness}\label{complete}

\begin{figure*}
\begin{center}
\epsfig{figure=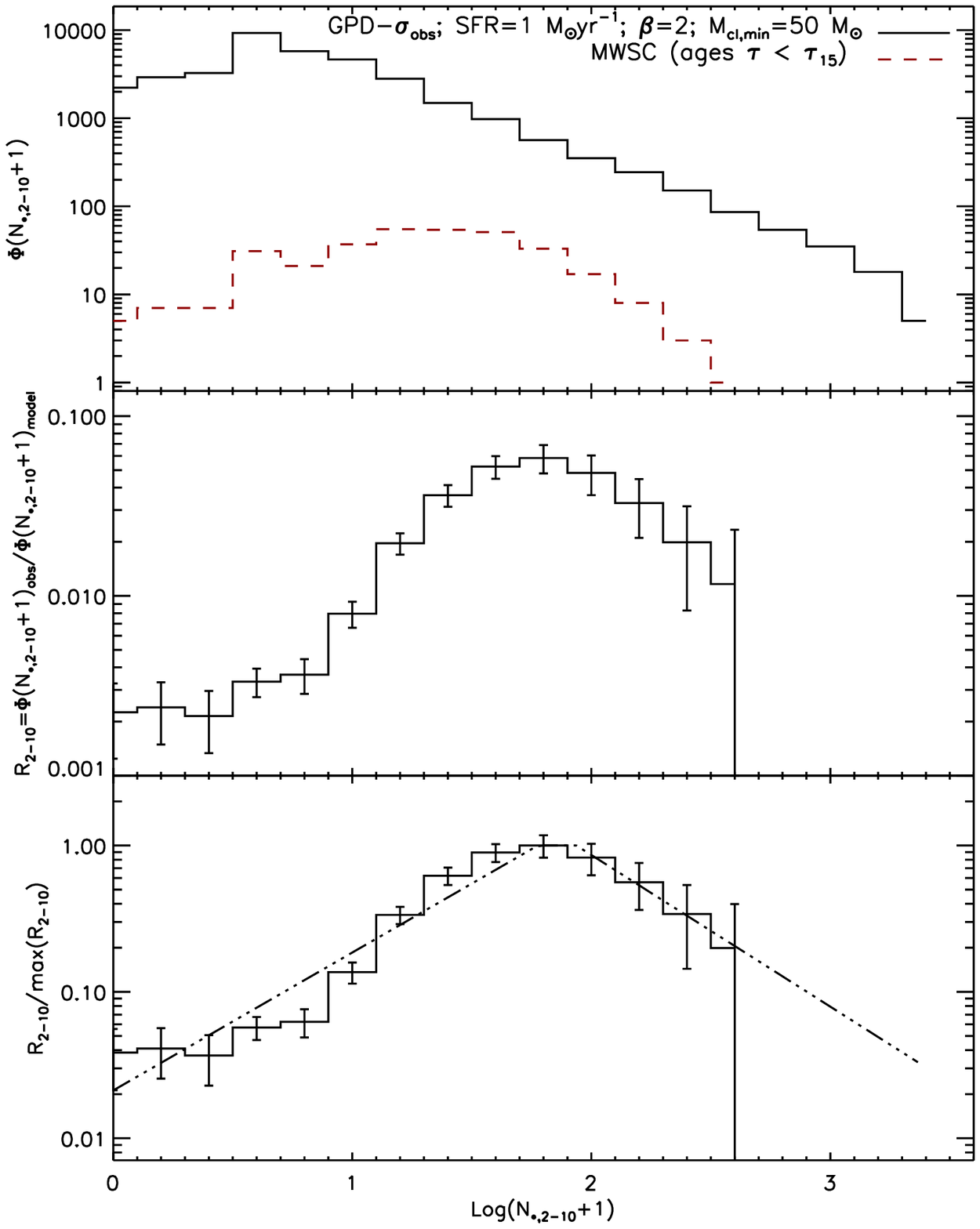,width=\columnwidth}
\hspace{0.5cm}
\epsfig{figure=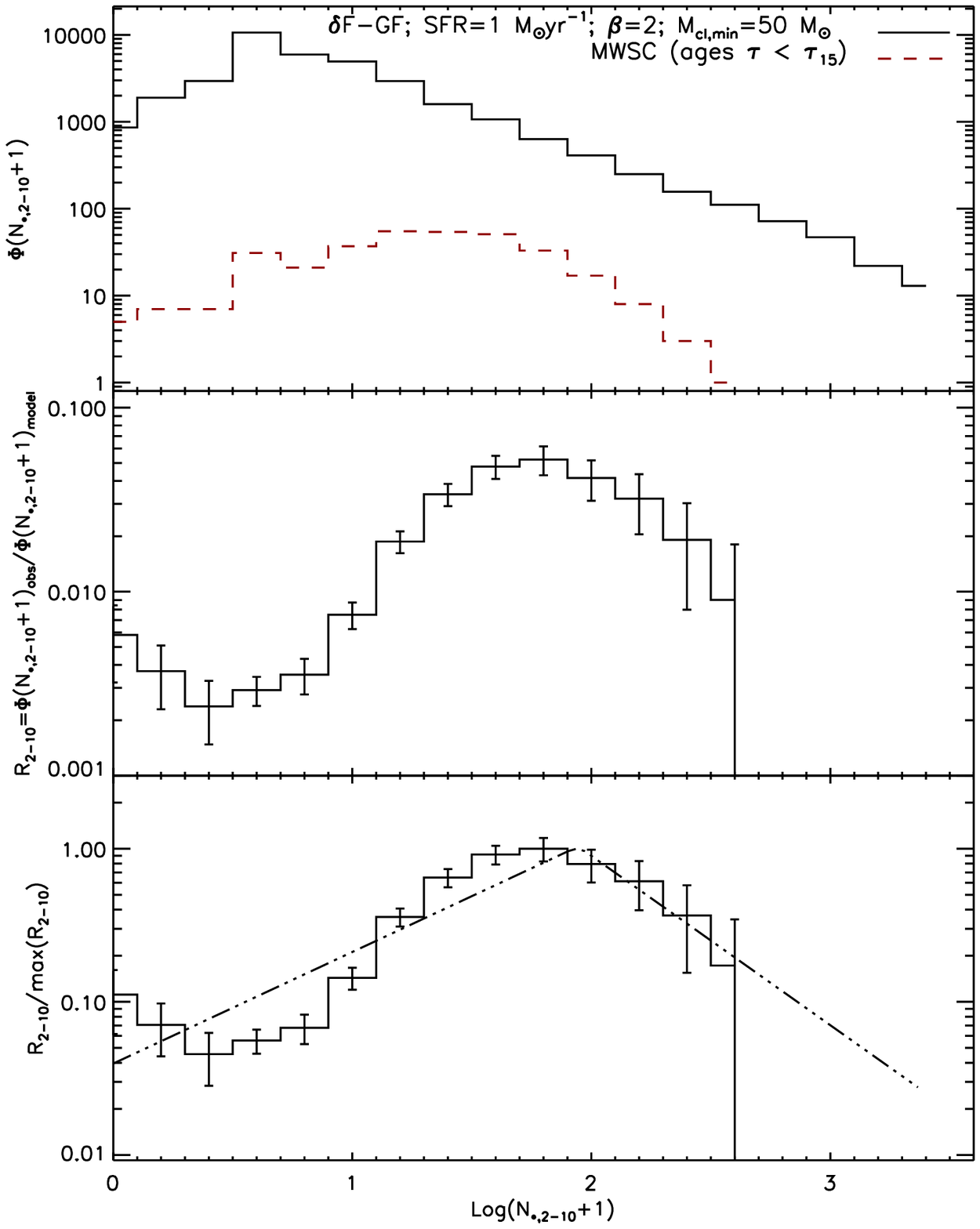,width=\columnwidth}
\end{center}
\vspace{0.5cm}
\caption{The completeness correction for two realizations of the ICLMF (left, a GPD-$\sigma_{obs}$ case, and, right, a $\delta$f-GF case). The other parameters are similar are similar in the two models, namely, SFR$=1$ M$_{\odot}$ yr$^{-1}$, $\beta=2$, and $M_{cl,min}=50$ M$_{\odot}$. The completeness correction is based on the clusterÕs low mass B star systems population (star-systems with masses between $2 {\rm M}_{\odot} \leq M_{*} \leq 10 {\rm M}_{\odot}$). The top panels display the distribution of the total number of low mass B stars/systems ($N_{*,2-10})$ for these realizations of the ICLMF (black full line) and for the ensemble of young clusters in the MWSC catalog (dashed purple line). The middle panels displays the ratio ($R_{2-10}$) of $N_{*,2-10}$ in the observations to the model. The lower panels display the ratio of $R_{2-10}$ normalized to its value at the peak. The quantity $f_{comp}=R_{2-10}/ R_{2-10}(peak)$ defines the completeness. The completeness function is approximated by a linear on both sides from the position of the peak. The fit to the completeness function is shown for these two examples with the triple dot-dash lines.}
\label{fig6}
\end{figure*}

Before we can compare the models to the observations, it is necessary that each cluster mass function, after the effects of the binary population and stellar evolution are taken into account is also corrected for effects of sample incompleteness. The MWSC sample is affected by incompleteness issues arising from the non-detection of clusters located at large distances from the Sun (both low and high-mass clusters) as well as the non-detection of faint nearby clusters.  Because massive clusters are scarce, they are less likely to be at close distances from the Sun and only a fraction of them will be detected. Furthermore, the census of low and intermediate mass stars even in detected clusters can be affected by effects of crowding.

The completeness correction is calculated for each realization of the cluster mass function with respect to the sample of young clusters in the MWSC survey. In this work, the approach we use in order to account for the effect of cluster incompleteness is based on the populations of low mass B stars in the clusters (i.e., stars with masses between $2 {\rm M}_{\odot} \leq M_{*} \leq 10 {\rm M}_{\odot}$) whose total number in a cluster is $N_{*,2-10}$. For each realization of the ICLMF, after the corrections for the effects of the binary population and stellar evolution have been taken into account, we compute the distribution of low mass B stars $\phi(N_{*,2-10})$. Fig.~\ref{fig6} (top panel, left) displays an example of the distribution of $\phi$ for one of the ICLMF realizations from the GPD-$\sigma_{obs}$ family of models (full black line) and for the young clusters in the MWSC catalog (purple dashed line) plotted versus ${\rm log}(N_{*,2-10}+1$). The unnormalized ratio of these two distributions ($R_{2-10}=\phi_{obs}/\phi_{model}$) is plotted in Fig.~\ref{fig6} (middle panel, left). A similar example is shown in the right panel for one realization of the ICLMF from the $\delta$f-GF family of models. All completeness functions that we have computed for the different synthetic cluster mass functions display a similar behavior, namely a peak at ${\rm Log}N_{*,2-10}\approx 1.8-2$, with a decrease for both increasing and decreasing values of  $N_{*,2-10}$ around the peak. We assume a completeness of unity at the position of the peak by normalizing the completeness function by its value at the peak, and fit linear relations for both components of the completeness function on each side of the peak. (triple dot-dash line, Fig.~\ref{fig6}, lower panel, left and right). As stated above, we interpret the decrease in completeness at low values of $N_{*2-10}$ by the non-detection of faint clusters, whereas the non-detection of some of the most massive clusters is most likely due to their relative scarcity in the Galaxy and the fact that they lie, on average, at larger distances from the Sun. The normalized function $f_{comp}=R_{2-10}/{\rm max}(R_{2-10})$ constitutes the completeness function. A cluster is admitted for the comparison with the observational data if its completeness probability $f_{comp}$ is larger than a uniform random number drawn between 0 and 1. For the families of ICLMFs generated in this work, typically only about half of the clusters in the ICLMF pass the filter of the completeness function and are used in the comparison with the observational data. Additional examples of the completeness function for various values of $\beta$ and its effect on the derived values of $f_{O,single}$ and $f_{O,lonely}$ are discussed in App.~\ref{appendixb}.

\section{Comparison of models to observations}\label{comparison}

\subsection{Models based on stochastic star formation}\label{stochastic}

\begin{figure}
\begin{center}
\epsfig{figure=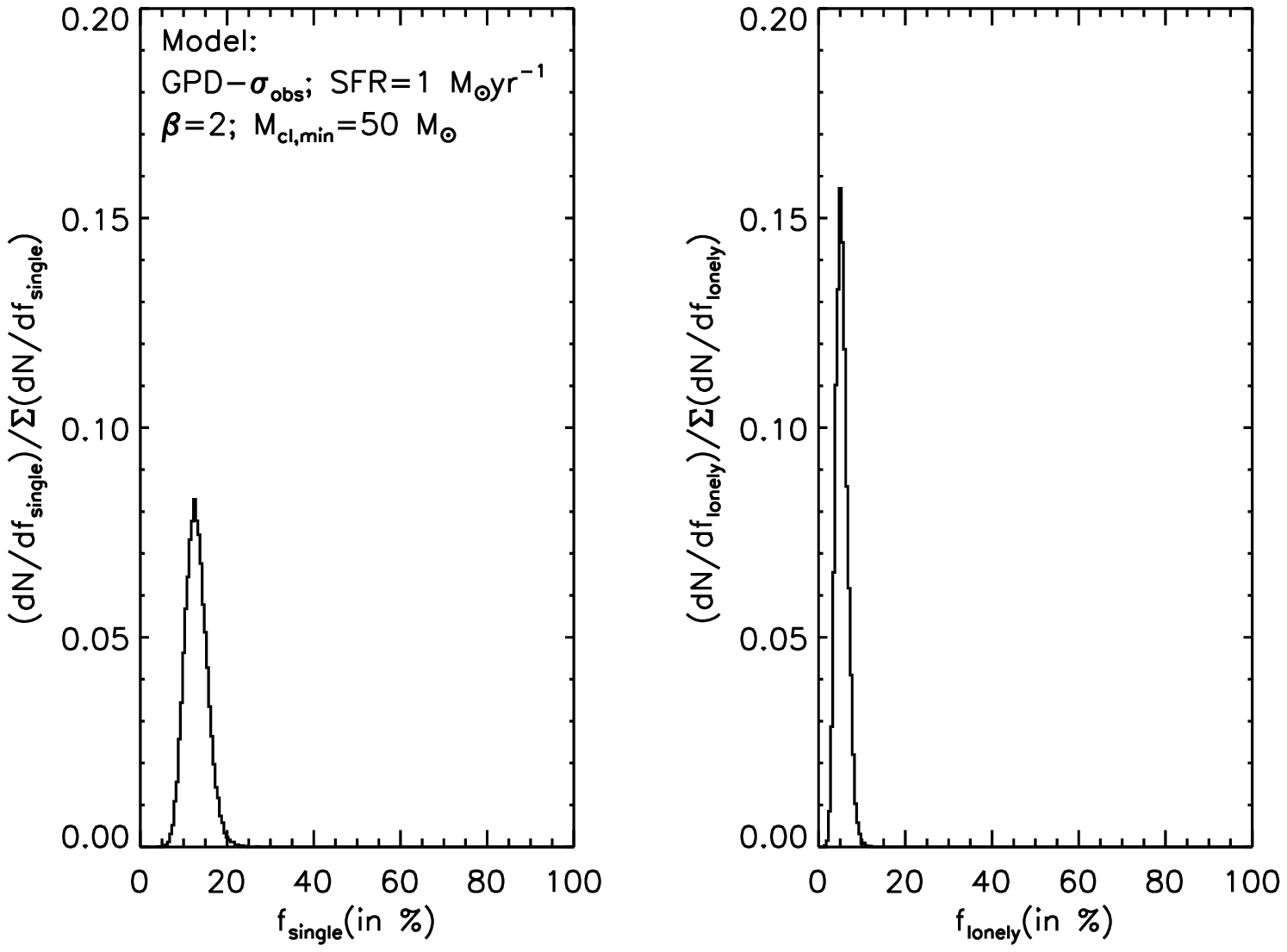,width=0.475\textwidth}
\epsfig{figure=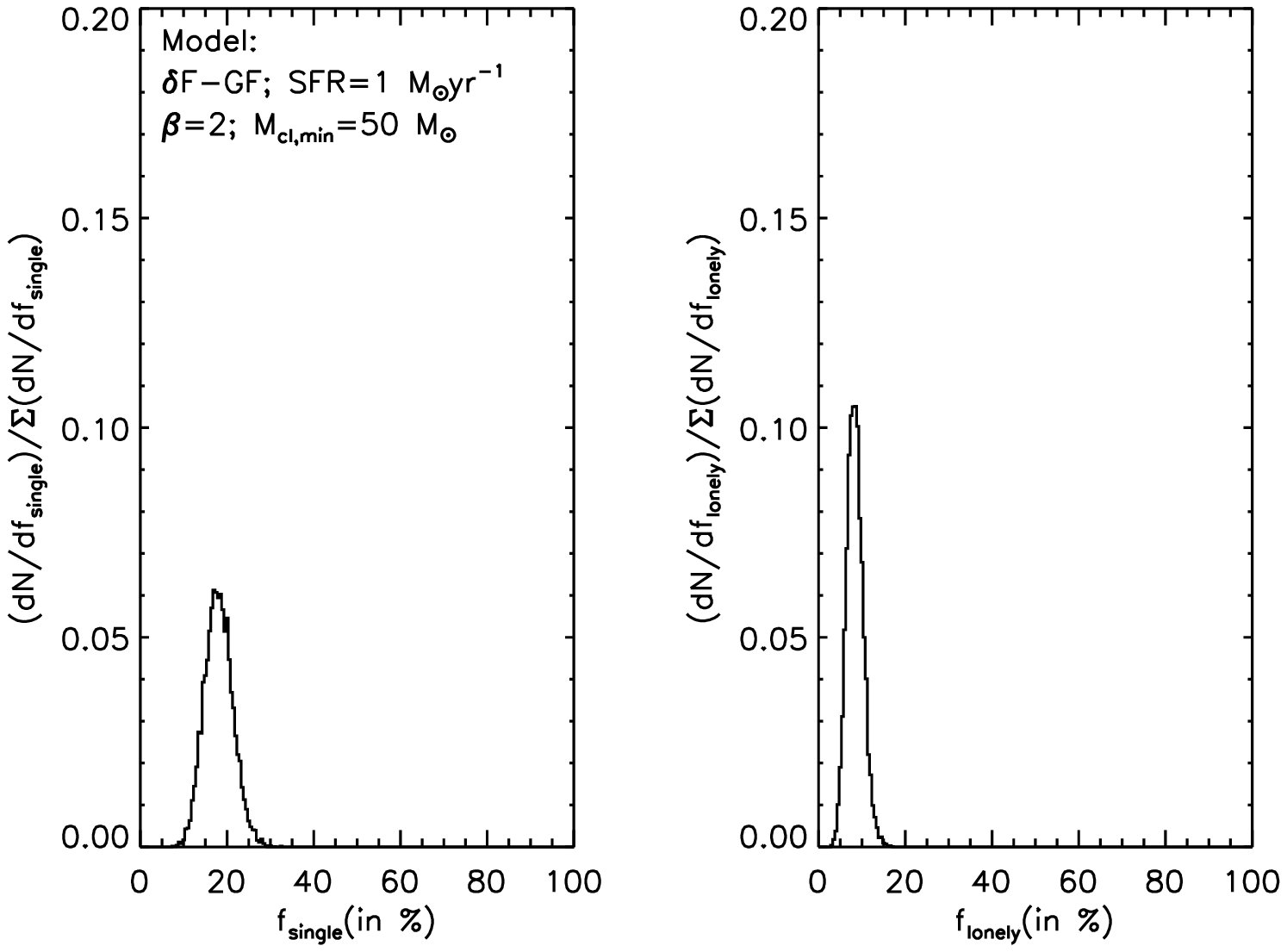,width=0.475\textwidth}
\end{center}
\caption{The probability distributions function of $f_{O,single}$ and $f_{O,lonely}$ in two realizations of the ICLMF (a GPD-$\sigma_{obs}$ case, top, and a $\delta$F-GF case, bottom). In both examples displayed here, the other parameters are similar SFR$=1$ M$_{\odot}$ yr$^{-1}$, $\beta=2$, and $M_{cl,min}=50$ M$_{\odot}$. The probability distributions of $f_{O,single}$ and $f_{O,lonely}$ are generated using measurements of these two quantities from 10000 randomly chosen subsamples of clusters, each of size equal to the size of young clusters in the MWSC catalog ($N_{MWSC,cl}=341$). Each subsample of clusters of size $N_{MWSC,cl}$ is drawn from the larger sample of clusters that pass the filter of the completeness function in a given realization of the ICLMF. The distributions are normalized by the total number of subsamples (i.e., 10000).}
\label{fig7}
\end{figure}

In this family of models, stellar masses in each cluster are randomly sampled in the mass range $[0.02, {\rm min}(M_{cl}, 150$)]M$_{\odot}$ for the set of parameters ($\Gamma$, $M_{ch}$, $\gamma$) that are assigned to the cluster. When comparing the fractions of single and lonely O stars ($f_{O,single}$ and $f_{O,lonely}$) between the observations and the models of synthetic clusters, we have two main options: option 1) we compare the observational values of $f_{O,single}$ and $f_{O,lonely}$ with the same quantities derived for the entire sample of clusters in a given initial cluster mass function that pass the filter of the completeness function (or to an average value of these quantities for a number of realizations of the ICLMF; i.e., the "all clusters" approach). Under this approach, the assumption is that the sample of $N_{\rm MWSC,cl}=341$ young clusters in the MWSC that are used to determine the observational values of $f_{O,single}$ and $f_{O,lonely}$ represents an unbiased sample of the Galactic population of young stellar clusters. For each family of IMF models, we perform 27 realizations of the ICMLF with different  permutations of the SFR, the randomly drawn ages of the clusters, and seed numbers used to randomly sample the ICLMF and the stellar masses within each cluster and measure the mean value and dispersion around the mean of $f_{O,single}$ and $f_{O,lonely}$ using these 27 realizations. A more accurate approach is achieved by comparing the observational values of $f_{O,single}$ and $f_{O,lonely}$ with the same quantities calculated from subsamples of synthetic clusters of size $N_{\rm MWSC,cl}=341$ (i.e., the subsamples approach). For each realization of the ICLMF, the subsamples are randomly drawn from the (much) larger sample of synthetic clusters that pass the filter of the completeness function. For each realization of the ICLMF, we randomly select 10000 subsamples of size $N_{\rm MWSC,cl}$ from the sample of clusters in the ICLMF that are accepted after passing the completeness correction, and for each subsample, we calculate The values of $f_{O,single}$ and $f_{O,lonely}$. Fig.~\ref{fig7} displays the probability distributions of $f_{O,single}$ and $f_{O,lonely}$ for two realizations of the ICLMF (top, for a GPD-$\sigma_{obs}$ case, and bottom for a $\delta$f-GF case). For both realizations displayed in Fig.~\ref{fig7} the other parameters are similar, namely, ${\rm SFR}=1$ M$_{\odot}$ yr$^{-1}$, $\beta=2$, and $M_{cl,min}=50$ M$_{\odot}$. Under this approach, for each realization, we measure the value of $f_{O,single}$ and $f_{O,lonely}$ as being the mean value from the 10000 subsamples and evaluate the corresponding dispersion. The mean value and mean dispersions for each family of models in then calculated as being the grand mean and mean dispersions of the 27 realizations of the ICLMF for each model. 

\begin{figure}
\begin{center}
\epsfig{file=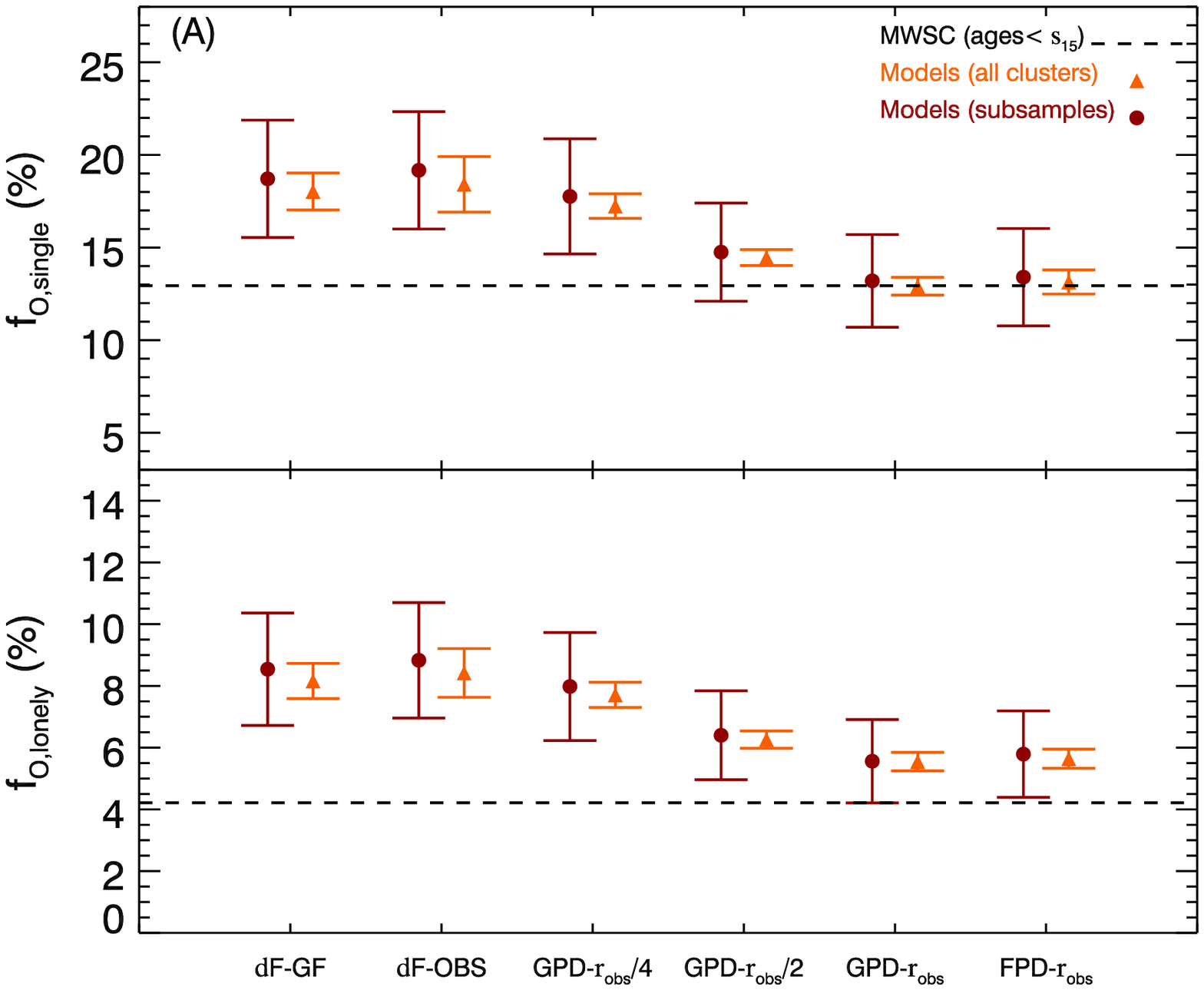,width=0.9\columnwidth} \\
\vspace{1cm}
\epsfig{file=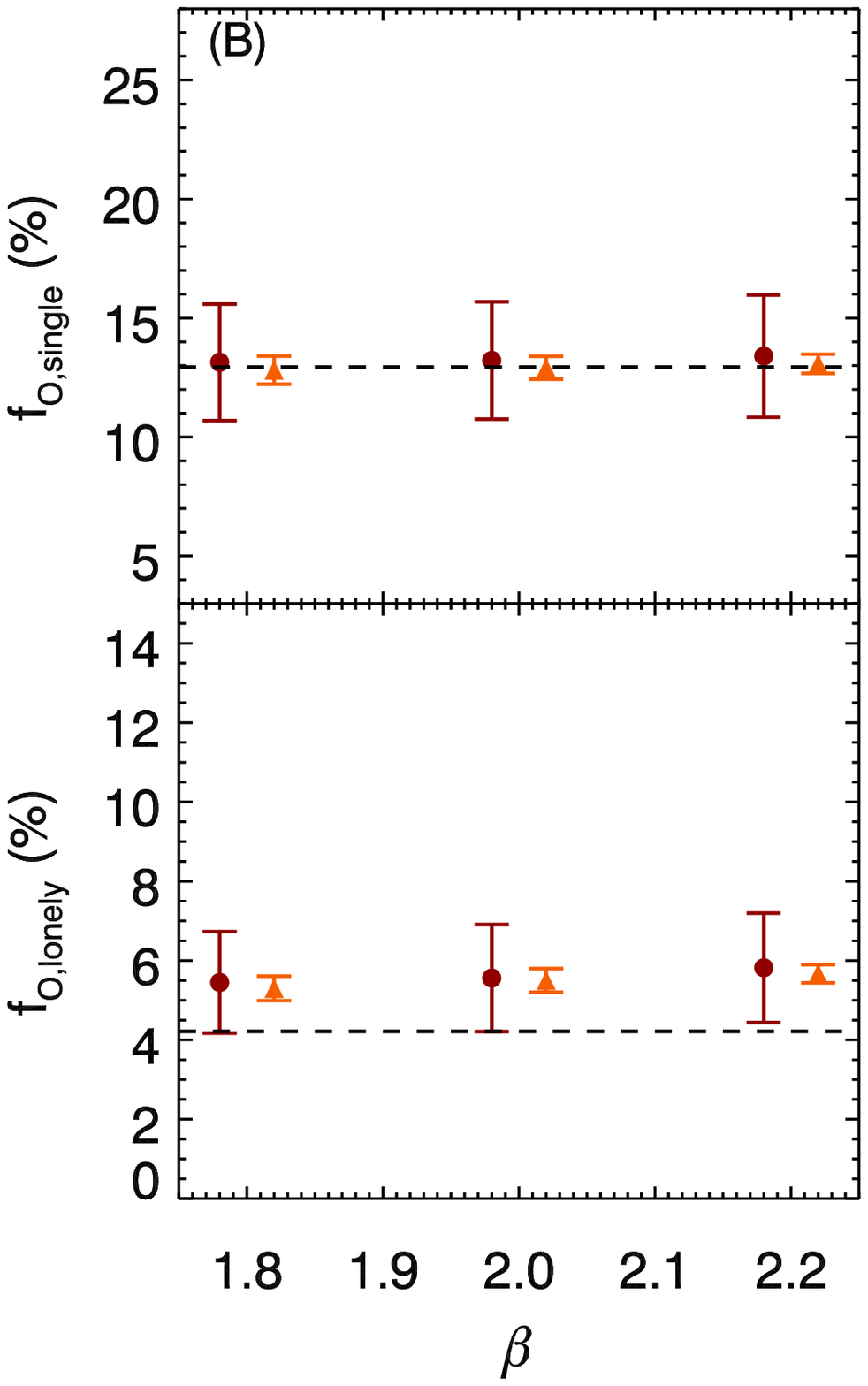,width=0.4\columnwidth} 
\hspace{0.1\columnwidth}
\epsfig{file=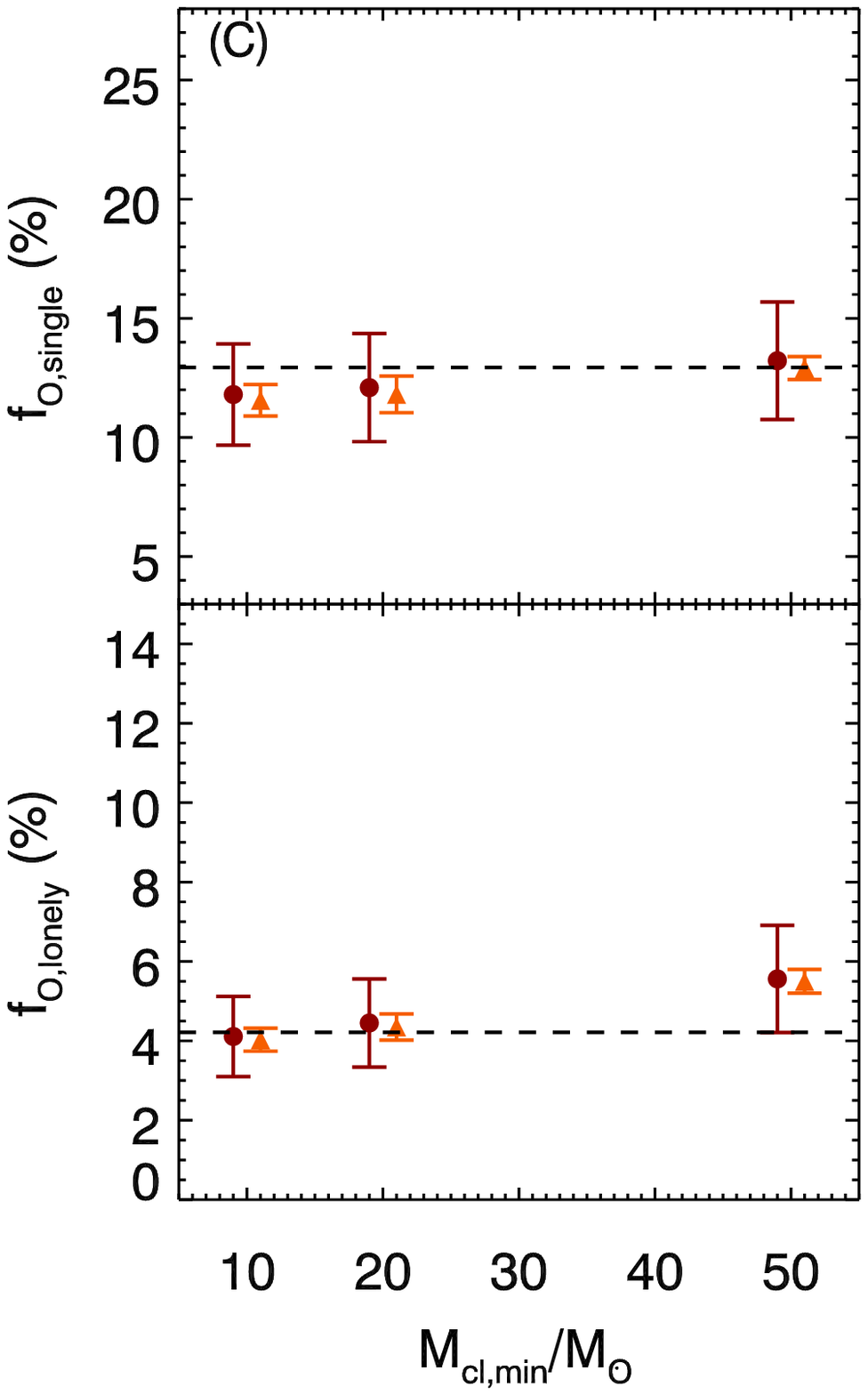,width=0.4\columnwidth} 
\vspace{1cm}
\caption{The top panel (A) in the figure compares the fractions of single and lonely O-star systems (defined as $M_{*} \geq 15$ M$_{\odot}$) calculated from the samples of clusters from the MWSC survey (dashed lines) with those measured for the different models of synthetic clusters (purple circles and orange triangles). The other parameters of the models are set at their fiducial values of $M_{cl,min}=50$ M$_{\odot}$, $M_{*,min=0.02}$ M$_{\odot}$, $M_{*,max}=150$ M$_{\odot}$, and $\beta=2$. The orange points and error bars are the mean and standard deviation of $f_{O,single}$ and $f_{O,lonely}$ calculated using all clusters that pass the filter of the completeness function in 27 realizations of the initial cluster mass function (ICLMF). The 27 realizations include variations of the Galactic star formation rates, the randomly drawn ages of the clusters, and seed numbers used to randomly sample the ICLMF and the stellar masses within each cluster. The purple points and error bars are the grand mean and grand mean absolute deviation from the 27 ICLMF realizations and where the mean and mean absolute deviation for each realization are calculated from 10000 drawings of subsamples of clusters each of size $N_{MWSC,cl}=341$. Each subsample of $N_{MWSC,cl}$ clusters is randomly drawn from the ensemble of clusters that pass the filter of the completeness function in each realization of the ICLMF. The lower left panel (B) and lower right panel (C) display the effect of changing the exponent of the ICLMF ($\beta$) and the minimum cluster mass ($M_{cl,min}$), respectively. For clarity, the orange triangles and purple circles have been shifted horizontally by [$0.2,-0.2$], and by [$1,-1$] in panels (B) and (C), respectively.}
\label{fig8}
\end{center}
\end{figure}

In Fig.~\ref{fig8} (panel A), we compare the values of $f_{O,single}$ and $f_{O,lonely}$ in the MWSC to those derived from the models of synthetic clusters generated with the various prescriptions for the distributions of the three IMF parameters. Each estimate of $f_{O,single}$ and $f_{O,lonely}$ in the "all sample" approach (orange points in Fig.~\ref{fig8}) is a mean value over 27 realizations based on three different random seeds for filling the ICLMF and the corresponding IMFs of the clusters, three permutations of the value of the Galactic SFR, and three permutations for the randomly assigned ages of the clusters. The associated error bars are the dispersions around the mean values measured from these 27 realizations. The purple points and associated error bars are, as described above, the grand mean and mean dispersion from the 27 ICLMF realizations and where the mean and dispersion for each realization are calculated from the 10000 drawings of subsamples of clusters each of size $N_{\rm MWSC,cl}=341$. 

As can be observed in Fig.~\ref{fig8} (panel A), a better agreement between the observations and the simulated clusters is achieved when the distributions of the IMF parameters are Gaussian functions that have significant intrinsic widths that are close to $\approx \left(\sigma_{\Gamma_{obs}},\sigma_{M_{ch,obs}}, \sigma_{\gamma_{obs}}\right)$. The equally good agreement between the FPD model and the observations shows that the results are not extremely sensitive to the exact shape of the distribution functions of the parameters. Future larger data sets of Galactic clusters will help better constrain the exact shape of the distribution functions of the parameters. Fig.~\ref{fig8} (panel A) also shows that, at more than the $2-\sigma$ confidence limit, no agreement is found between the observations and the models with narrow distributions of the IMF parameters and in particular when the distributions are delta functions located at either the values of the parameters for the Galactic field or at the mean values derived for the sample of young clusters by Dib (2014a). For realizations of the ICLMF in the GPD-$\sigma_{obs}$ family of models, between $42.2 \%$ and $69.1\%$ of the values of $f_{O,single}$ from the 10000 subsample realizations lie above the observational value (with an average of $\approx 55.5\%$ for the 27 realizations), and between $74.9\%$ and $94.4\%$ in the case of $f_{O,lonely}$ (with an average of $\approx 87.2\%$). In contrast, in the case of realizations of the ICLMF with the $\delta f$-GF family of IMF models, between $94.4\%$ and $99.1\%$ of the realizations of $f_{O,single}$ lie above the observational value (with an average of $\approx 97.1 \%$ for the 27 realizations), and between $99.3\%$ and $99.97\%$ of the values of $f_{O,lonely}$ lie above the observational value (with an average of $\approx 99.7\%$)

We also explore the effects of varying the exponent of the ICLMF and of the lower cluster mass cutoff for fixed values of the width of the Gaussian distributions (fixed at $\sigma_{\Gamma_{obs}}$, $\sigma_{M_{ch,obs}}$, and $\sigma_{\gamma_{obs}}$, for the distributions of $\Gamma$, $M_{ch}$, and $\gamma$ respectively). We show in Fig.~\ref{fig8} (panel B) the predicted single and lonely O star fractions for three values of $\beta=1.8, 2$, and $2.2$. In principle, a change in the value of $\beta$ strongly affects the relative fraction of low mass- to massive clusters, which translates into significant variations in the fraction of single and lonely O-stars. Steeper/shallower values of $\beta$ result in a larger/smaller fraction of low mass clusters which are more/less likely to harbor single and lonely O stars. However, a significant fractions of these variations are "washed away" by the completeness function. The completeness function for various values of $\beta$ and its effect on the derived values of $f_{O,single}$ and $f_{O,lonely}$ are discussed in more detail in App.~\ref{appendixb}. 

As such, the comparisons in Fig.~\ref{fig8} (panel B) do not particularly constrain the value of $\beta$. It shows, however, that the broad distributions of the IMF parameters are required in order to better reproduce the observational values of $f_{O,single}$ and $f_{O,lonely}$, regardless of the value of $\beta$. The effect of changing the lower mass cutoff of the ICLMF ($M_{cl,min}$) is displayed in Fig.~\ref{fig8} (panel C). The comparison when $M_{cl,min}$ is 10, 20 and 50 M$_{\odot}$ shows that the fractions of single and lonely O stars are not extremely sensitive to the value of the lower mass cutoff. This can be easily understood as due to the fact that low mass clusters below $50$ M$_{\odot}$ seldom contain any O star. We note, however, that the best agreement between the models and the observations is for $M_{cl,min}=10$ M$_{\odot}$. This implies that the ICLMF may well extend to masses close to 10 M$_{\odot}$.

Stars in clusters interact dynamically, and three-body interactions can lead to the ejection of stars or binary systems. The fraction of O stars that are ejected can vary widely from cluster to cluster, and can depend on the mass of the cluster, its half mass radius, the degree of primordial mass segregation of stars in the cluster, the binary fraction, mass ratios in massive binaries, and the period distributions of massive systems. For the most realistic estimates of these parameters, the fraction of ejected O stars has been estimated to be negligible for clusters whose masses are $\lesssim 400-500$ M$_{\odot}$. It increases up to $\approx 25 \%$ for cluster masses of $\approx 3000$ M$_{\odot}$ and declines to $5-10 \%$ for higher cluster masses (Oh et al. 2015). Note that these fractions were evaluated for clusters that obey an $M_{cl}-M_{*,max}$ relation and all possess the same underlying Galactic field-like IMF (i.e., the Kroupa IMF). As such, these estimates may not reflect exactly the expected fractions of ejected O stars for each family of our synthetic clusters. However, the basic result, that dynamical ejection of O stars from clusters less massive than $\approx 400-500$ M$_{\odot}$ is insignificant, should not depend on these details. Since most clusters that harbor single O stars in our models have masses $\lesssim 400-500$ M$_{\odot}$, the ejection of massive stars from the clusters is not expected to significantly affect the value of $f_{O,single}$. If we account for the fraction of ejected O stars, The quantity $f_{O,single}$ can be approximated by $\approx N_{O,single}/ (N_{O,single}+f_{O,ejec} N_{O,non-single}$), where $N_{O,non-single}$ is the number of O star that are not single, and $f_{O,ejec}$ is the fraction of ejected O stars as a function of the cluster mass. Using the values of $N_{O,single}$ and $N_{O,non-single}$ and adapted values of $f_{O,ejec}$ as a function of cluster mass for the different families of models yields small increases in $f_{O,single}$ of $\approx 10-12 \%$ for GPD-$\sigma_{obs}$ type models and $\approx 18-22 \%$ for $\delta$F-GF type models. Applying this correction to account for the fraction of dynamically ejected O stars increases the disagreement between the $\delta$F-GF type models and the observations while at the same time, it does not substantially affect the relative good agreement between the GPD-$\sigma_{obs}$ models and the observations.

\subsection{Models with an imposed $M_{cl}-M_{*,max}$ relation}\label{mclmmax}

Our modeling allows us to test the consequences of constraining the maximum stellar mass in clusters. Vanbeveren (1982) and Weidner \& Kroupa (2004) argued that a deterministic relation exists between the mass of the most massive star in a cluster and the mass of the cluster. The existence of a cluster-mass-dependent truncation of the IMF is highly debated and has important consequences for cluster and galaxy properties and evolution. A $M_{cl}-M_{*,max}$ relation results in a steeper galaxy wide IMF for lower mass galaxies at a fixed SFR and to a downturn in the ratio of the H$\alpha$ emission to the Far Ultraviolet emission (FUV) at low galactic FUV luminosities. A number of observational studies found that a cluster-mass-dependent truncation of the IMF leads to an under-prediction of the observed H$\alpha$ luminosities at low FUV luminosity (Fumagalli et al. 2011; Weisz et al. 2012). Other studies on the scale of resolved star forming regions using the H$\alpha$/FUV ratios or the correlation between H$\alpha$ and bolometric luminosities found results that do not seem to lend support to the existence of an $M_{cl}-M_{*,max}$ relation (Hermanowicz et al. 2013). 

\begin{figure}
\begin{center}
\epsfig{file=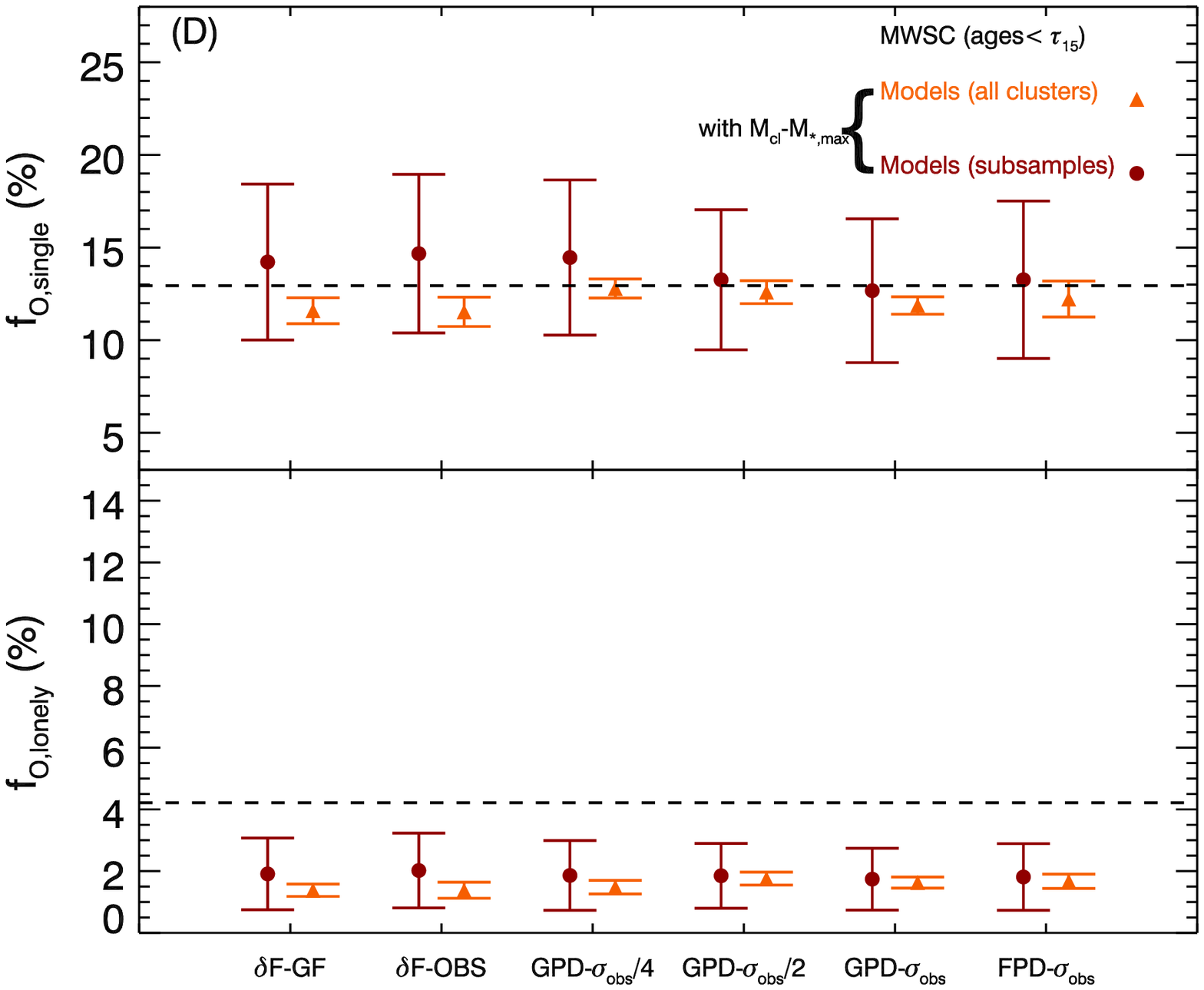,width=0.9\columnwidth} \\
\vspace{1cm}
\epsfig{file=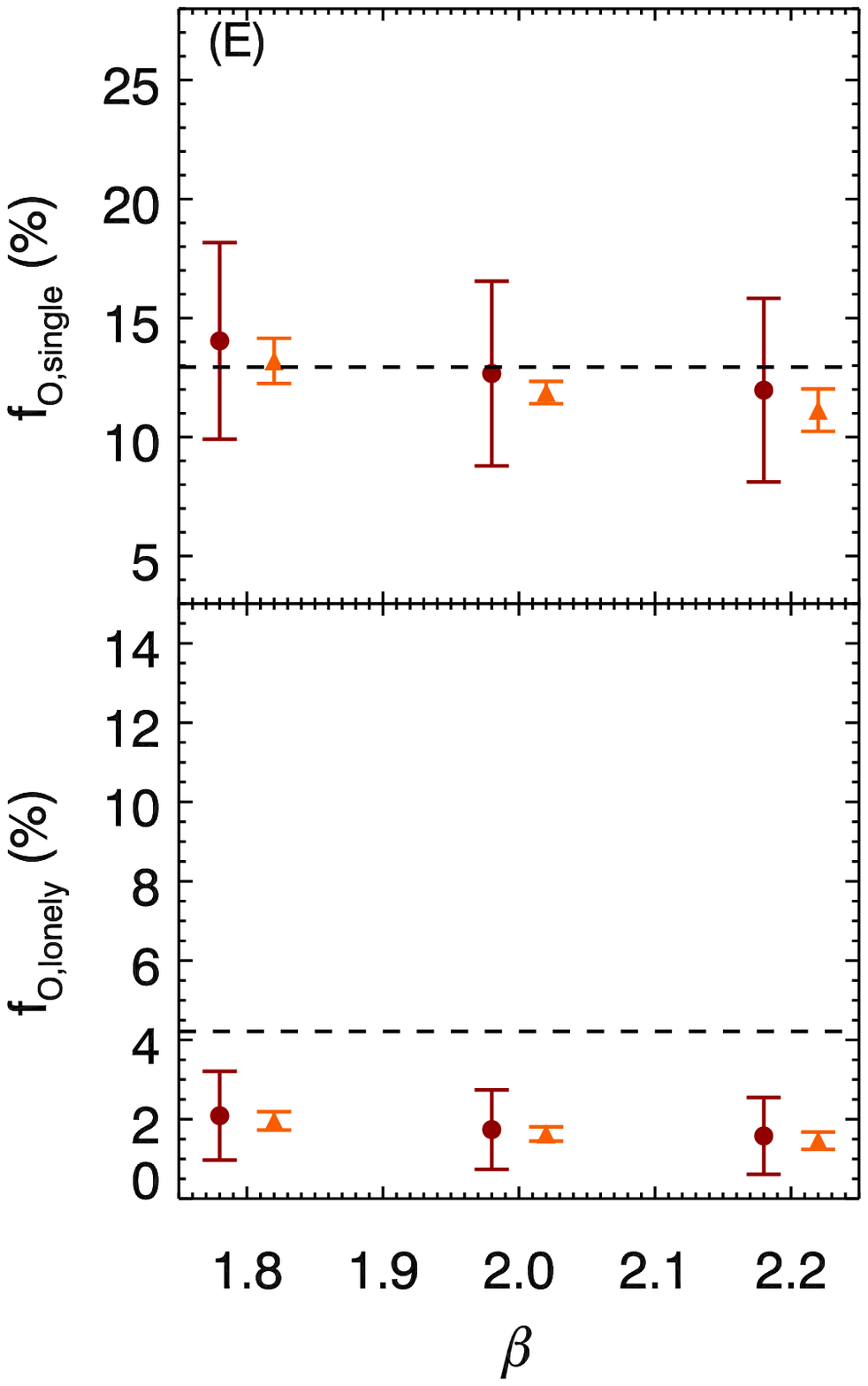,width=0.4\columnwidth} 
\hspace{0.1\columnwidth}
\epsfig{file=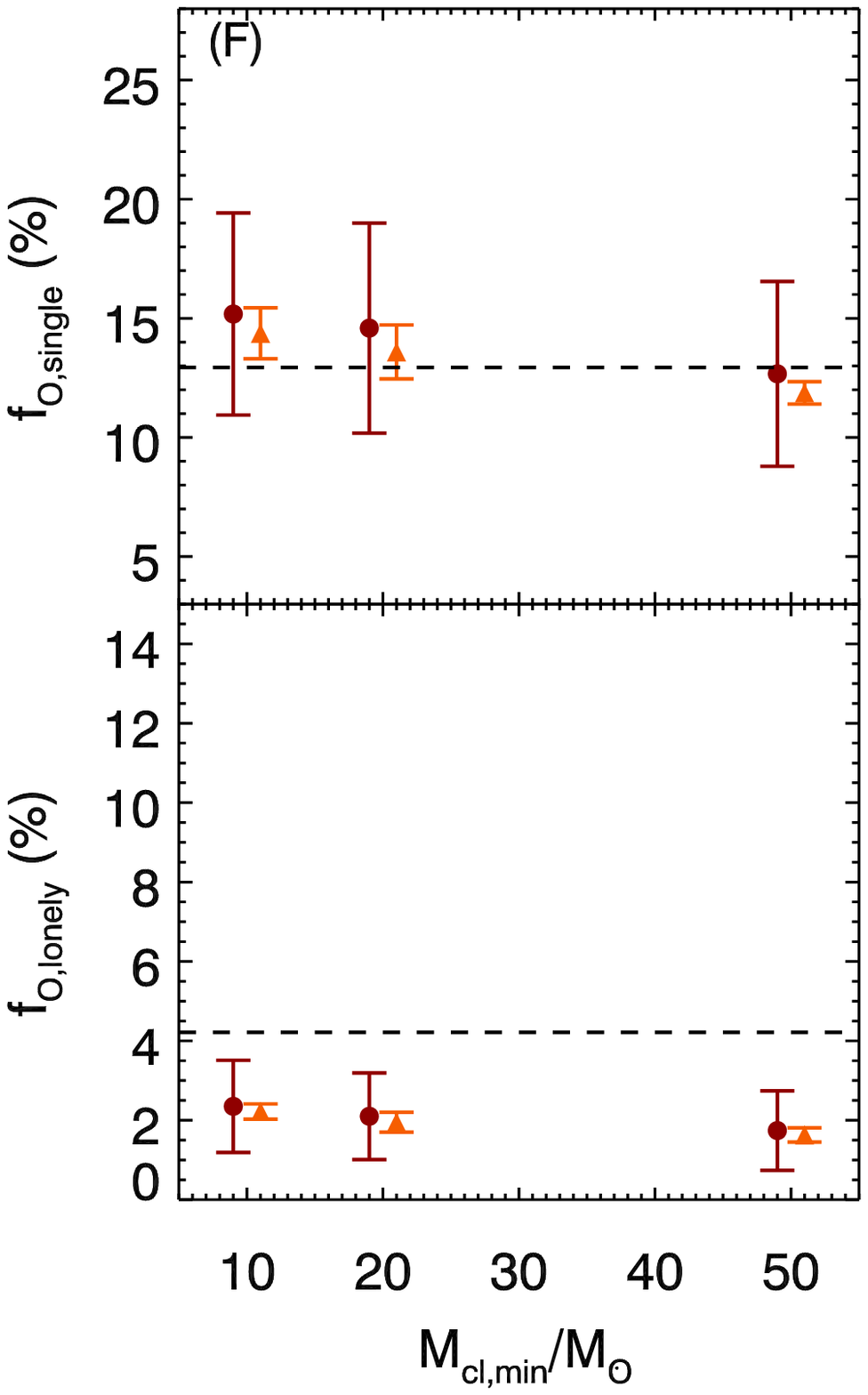,width=0.4\columnwidth} 
\vspace{1cm}
\caption{Effects of the cluster mass ($M_{cl}$)-maximum stellar mass ($M_{*,max}$) relation on the fraction of single and lonely O stars. This figure is similar to Fig.~\ref{fig8} with the exception that there is an imposed $M_{cl}-M_{*,max}$ relation which determines the maximum mass a star can have when sampling stellar masses in a cluster of mass $M_{cl}$ (see text for more details). For clarity, the orange triangles and purples circles have been shifted horizontally by [0.2,-0.2] and by [1,-1] in panels (E) and (F), respectively}
\label{fig9}
\end{center}
\end{figure}

Following the same procedure described above, we generate additional models in which the masses of stars (i.e., star-systems) in the clusters are randomly sampled in the range $M_{*,min}=0.02$ M$_{\odot}$ and an $M_{*,max}$ that is imposed by the latest version of the $M_{cl}-M_{*,max}$ relation (Weidner et al. 2013). The $M_{cl}-M_{*,max}$ relation is given by ${\rm log}_{10}(M_{*,max}/{\rm M}_{\odot})=-0.66+1.08\times \left[{\rm log}_{10}(M_{cl}/{\rm M}_{\odot})\right]-0.15\times\left[{\rm log}_{10}(M_{cl}/{\rm M}_{\odot})\right]^{2}+0.0084\times\left[{\rm log}_{10}(M_{cl}/{\rm M}_{\odot})\right]^{3}$ and is assumed to be valid for cluster masses $M_{cl} \leq 2.5 \times 10^{5}$ M$_{\odot}$ which is the case of the clusters considered in this work. All other parameters are kept at their fiducial values. We measure $f_{O,single}$ and $f_{O,lonely}$ in this additional set of models. The results displayed in Fig.~\ref{fig9} (Panel D) show that these models do not satisfactorily reproduce the observations. While the comparison using $f_{O,single}$ is inconclusive, imposing a $M_{cl}-M_{*,max}$ relation leads to an underestimate of $f_{O,lonely}$ by a factor of $\approx 2$, at the $1-\sigma$ confidence interval, for all models with respect to the observational value. This conclusion is not sensitive to the choice of $\beta$ and $M_{cl,min}$ (Fig.~\ref{fig9}, panels E and F, respectively). While our results do not entirely rule out the $M_{max}-M_{cl}$ proposed by Weidner et al. (2004,2013), they do cast serious doubts on its existence. 

\subsection{Additional predictions}

\begin{figure*}
\begin{center}
\epsfig{figure=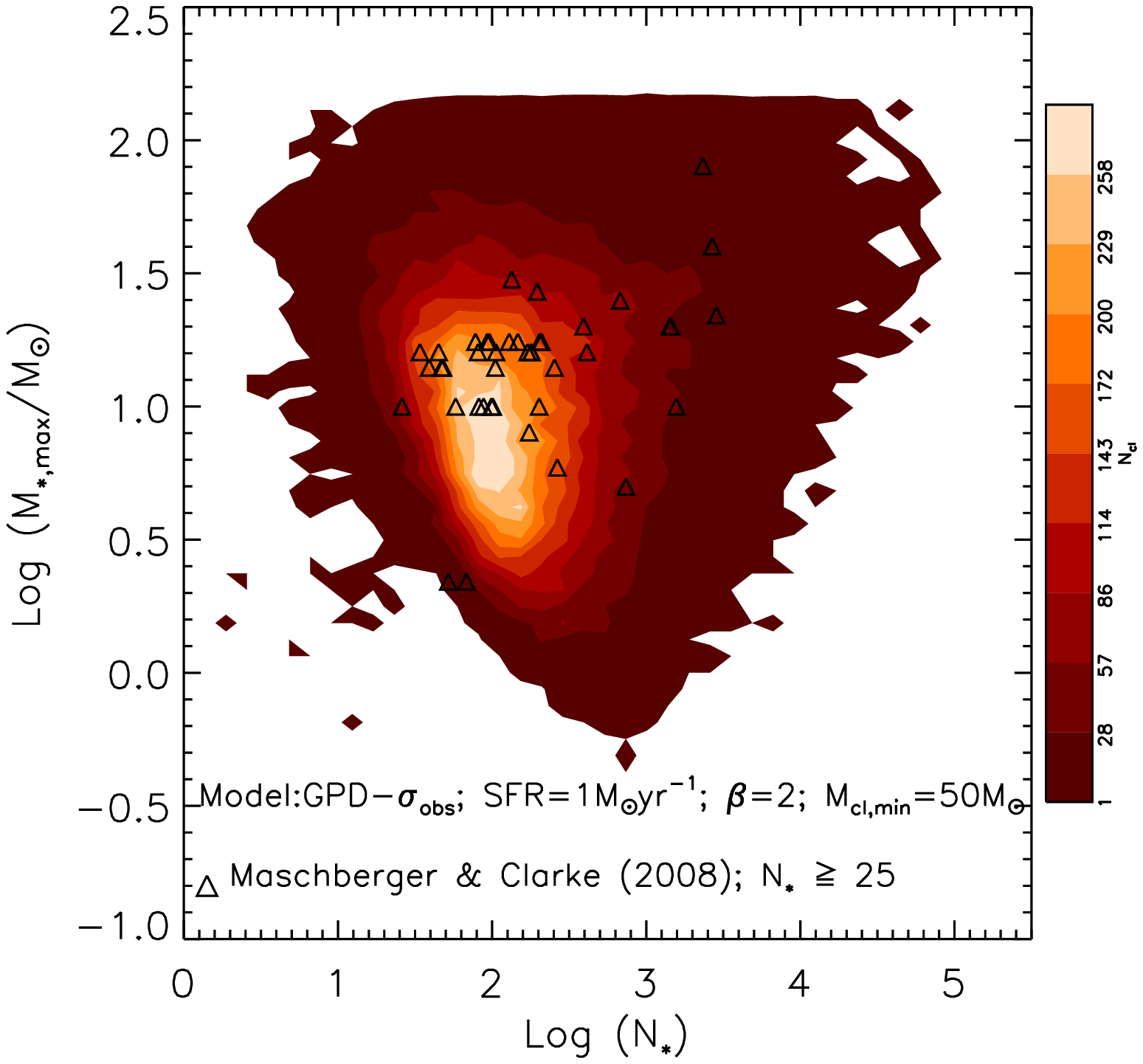,width=0.475\textwidth}
\epsfig{figure=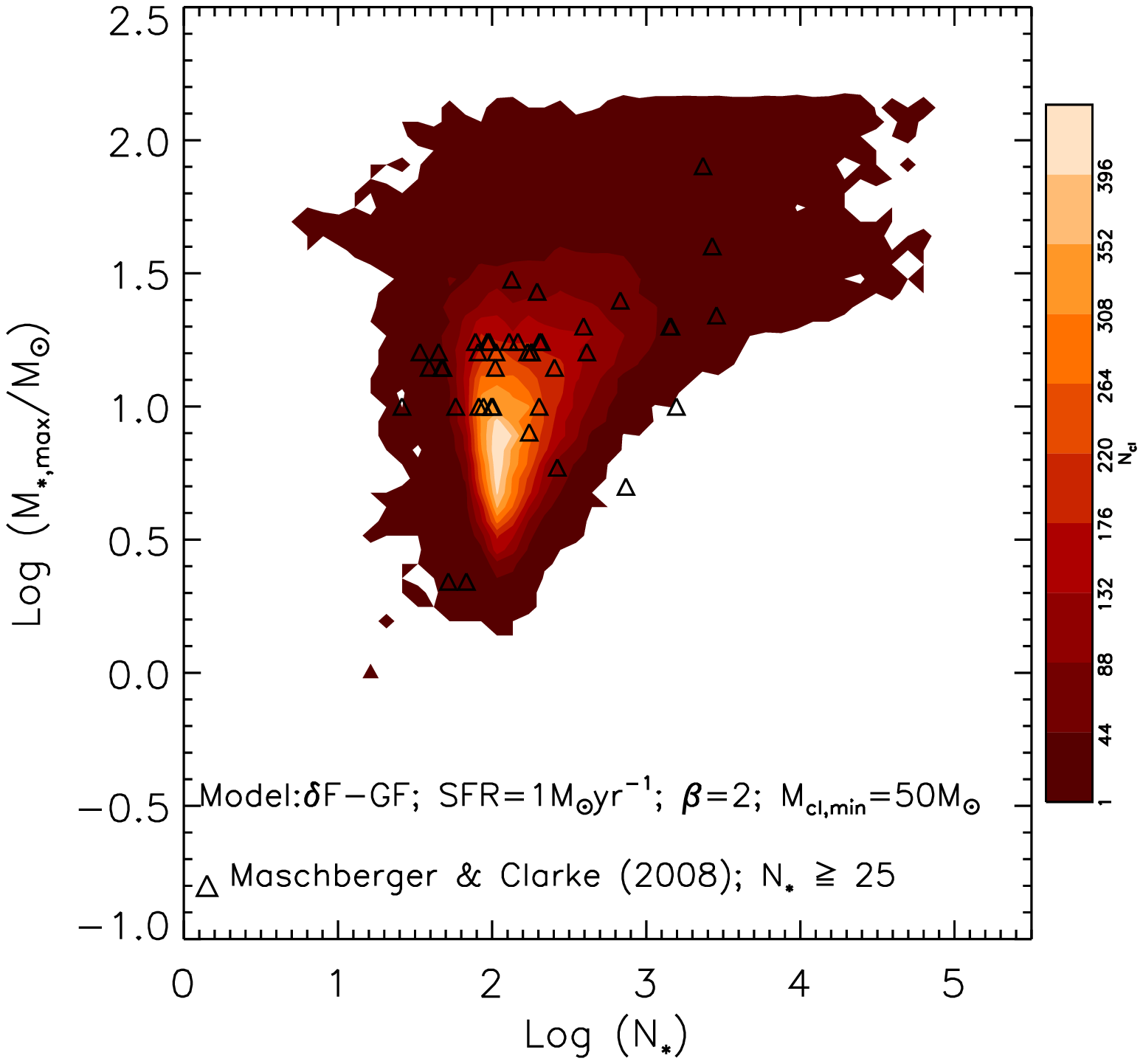,width=0.475\textwidth}
\end{center}
\caption{The $N_{*}-M_{*,max}$ relation. Comparison of the relationship between the number of stars in the clusters ($N_{*}$) and the mass of the most massive star in the clusters ($M_{*,max}$) in two realizations of the ICLMF. The left panel displays a case in which the set of three parameters that describe the IMF of each cluster are each randomly drawn from a GPD-$\sigma_{obs}$ probability distribution function whereas the right panel displays a case in which the set of three parameters that describe the IMF is similar to the values of the parameters for the Galactic field mass function. Overlaid are observational data compiled by Maschberger \& Clarke (2008).}
\label{fig10}
\end{figure*}

\begin{figure*}
\begin{center}
\epsfig{figure=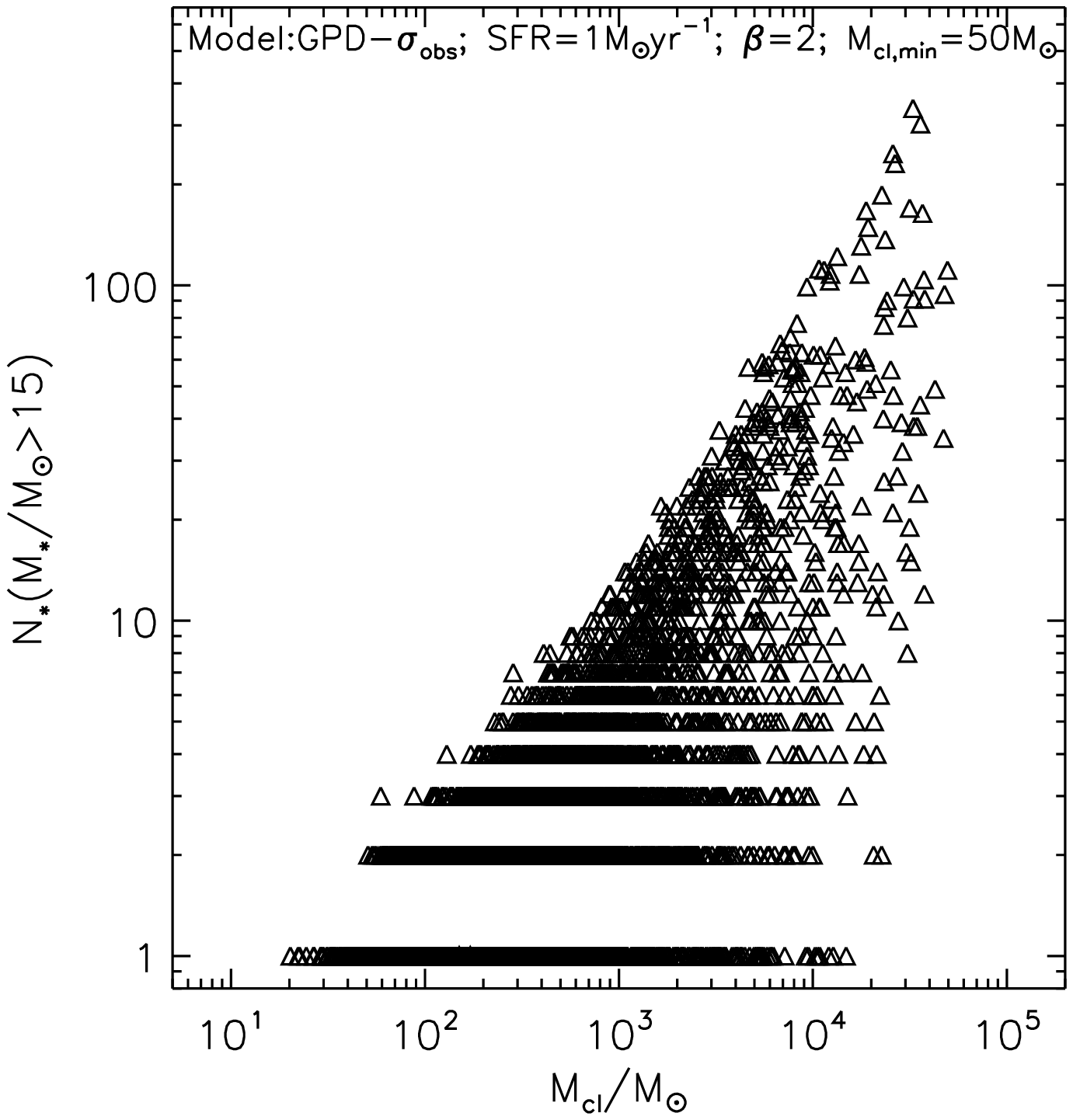,width=0.475\textwidth}
\epsfig{figure=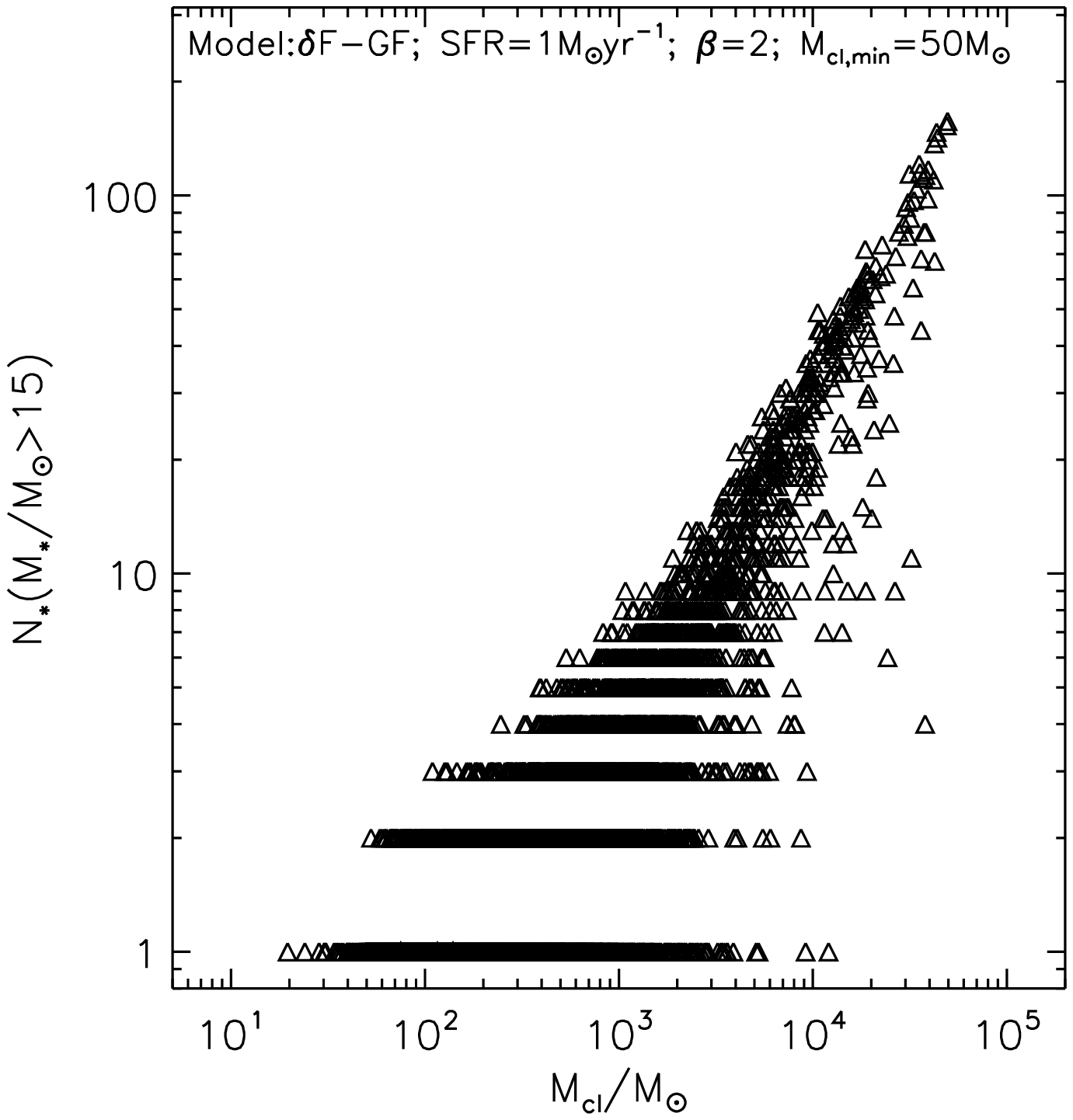,width=0.475\textwidth}
\end{center}
\caption{The $N_{*}(M_{*}/{\rm M}_{\odot}>15$)-$M_{cl}$ relation. Comparison of the relationship between the mass of the cluster ($M_{cl}$) and the number of stars more massive than 15 M$_{\odot}$ ($N_{*}(M_{*}/{\rm M}_{\odot}>15$)) present in each cluster in two realizations of the ICLMF. The left panel displays a case in which the set of three parameters that describe the IMF of each cluster are each randomly drawn from a GPD-$\sigma_{obs}$ probability distribution function whereas the right panel displays a case in which the set of three parameters that describe the IMF is similar to the values of the parameters for the Galactic field mass function.}
\label{fig11}
\end{figure*}

The primary goal of this work is to provide a method that allows us to assess the universality of the IMF for the population of Galactic stellar clusters across the entire stellar mass range and that is based solely on the clusters stellar populations of massive O stars (for discriminating between IMF models) and B stars (for accounting for the completeness effects). The method does not rely on the knowledge of the exact total number of stars ($N_{*}$) in the clusters, nor does it rely on the knowledge of the masses of the low mass stars (i.e., masses $M_{* }< 2$ M$_{\odot}$). Nevertheless, with this approach, it is possible to make a number of additional predictions for the models with the different families of the IMF parameters distribution functions. These predictions can be contrasted with additional observational constraints, when available. Fig.~\ref{fig10} displays a 2D histogram of the relationship between the numbers of stars found in clusters ($N_{*}$) and the maximum stellar mass in the clusters ($M_{max,*}$) for two realizations of the ICLMF. One of these realizations of the ICLMF uses GPD-$\sigma_{obs}$ distributions functions of the IMF parameters (left panel) and the other $\delta$F-GF distributions of IMF parameters (right panel). The results displayed in Fig.~\ref{fig10} suggest the existence of a different $N_{*}-M_{*,max}$ relation between these models. The observational data overlaid to the models in Fig.~\ref{fig10} comes from a compilation of young clusters by Maschberger \& Clarke (2008). We only include clusters with an observational value of $N_{*} \geq 25$. While the number of observed clusters in this compilation is relatively small compared to the number of clusters in each of the simulated models, we can tentatively argue that the observational data points in Fig.~\ref{fig10} lie closer to the peak of the 2D distributions when the IMF parameters are described by the GPD-$\sigma_{obs}$ case versus the $\delta$F-GF case. Fig.~\ref{fig11} displays the cluster mass ($M_{cl}$) - Number of O stars more massive than 15 M$_{\odot}$ ($N_{*}(M_{*}/{\rm M}_{\odot} >15$)) relation for the same two realizations with the GPD-$\sigma_{obs}$ and the $\delta$F-GF distribution functions of the IMF parameters. Here also, noticeable differences can be observed in this scatter relation. In particular, there is a much tighter correlation between $M_{cl}$ and $N_{*}(M_{*}/M_{\odot}>15$) for the case with a universal IMF ($\delta$F-GF), particularly at high cluster masses.

\section{Comparison to previous work}\label{otherwork}

To the best of our knowledge, this paper presents the first attempt to constrain the distribution functions of the set of parameters that describe the shape of the IMF over the entire stellar mass range for a large population of young clusters in the Milky Way. Models of synthetic clusters have been used by other authors in order to infer the fraction of single (or isolated\footnote{Both de Wit et al. (2007) and Parker \& Goodwin (2007) define an O star as a single star in a cluster with $M_{*} \geq 17.5$ M$_{\odot}$, and B stars as stars with 10 M$_{\odot}  \leq M_{*} \leq 17.5$ M$_{\odot}$}) O stars  and compare it to the putative fraction of isolated O star in the Galactic field (e.g., de Wit et al. 2005; Parker \& Goodwin 2007; Lamb et al. 2010; Weidner et al. 2013). In all of these models, however, zero-age age clusters were always constructed under the assumption of a universal IMF and most of them did not include additional corrections (binarity, stellar evolution, and incompleteness effects). Interestingly, Parker \& Goodwin (2007) found values of $f_{O,single}$ and $f_{O,lonely}$ of $16.7\%$ and $9.7\%$ when sampling the IMF in clusters stochastically, for a ICLMF with $\beta=2$, and using a Kroupa-like constant Galactic field IMF (see Table 1 in their paper). These values are in relative good agreement with the values we find using the $\delta$F-GF family of models (Fig.~\ref{fig8}), and both are higher than the corresponding values measured from the MWSC. They also found that these fractions are reduced by a factor of $\approx 2-3$ when stellar masses are randomly sampled under the constraint of an $M_{cl}-M_{*,max}$ relation.   

In term of methodology, our approach sits in between models of zero-age populations of clusters (e.g., Parker \& Goodwin 2007) and fully fledged population synthesis models which investigated the effects of stochasticity, shape of the ICLMF, binary fraction, and variations of the IMF on the distributions functions of cluster properties and on global galactic properties (e.g., Cervi\~{n}o \& Lurdiana 2006,2009; Conroy et al. 2009; Eldridge \& Stanway 2009; da Silva et al. 2012,2014; Cervi\~{n}o 2013)
   
\section{DISCUSSION}\label{discussion}

The debate over the universality or potential variation of the IMF among stellar clusters, as well as the similarity between the IMF in clusters and the Galactic field stellar mass function has been ongoing ever since Salpeter (1955) published his findings. From a theoretical point of view, much of the arguments in favor/disfavor of variations of the IMF originate from the inclusion/absence in the models of the necessary physical processes that can lead to a significant degree of variations. A perfect illustration of this are the contrasting conclusions made by Dib et al (2010) and Hennebelle (2012). Dib et al. (2010) considered the case of  accreting protostellar cores in a non-accreting star forming clump, whereas Hennebelle (2012) considered the case of non-accreting cores in an accreting clump. Dib et al. (2010) showed that the accretion of gas by protostellar cores can lead to variations in the core mass function (and hence of the IMF) when environmental conditions vary from clump-to-clump. Dib et al. (2010) and Dib (2014b) showed that a Taurus-like mass function can be reproduced when protostellar cores continue to accrete over longer timescales (i.e., as a result of being supported by stronger magnetic fields), and this leads to the depletion of the population of low mass cores and shifts the peak of the mass function towards higher masses. In contrast, in the model of Hennebelle (2012), the accretion of gas by the clump from the larger scale environment is only expected to change the thermodynamical properties of the gas out of which newer generations of stars can form in the clump. Hennebelle (2012) finds that the position of the peak of the IMF is not extremely sensitive to the thermodynamical conditions of the star forming gas, as earlier suggested by Elmegreen et al. (2008), and more recently confirmed by Krumholz et al. (2016). 

Several observational studies have also reported that the slope of the IMF at the high mass end of starburst clusters such as the Arches cluster, NGC 3603, and the Quintuplet cluster might be shallower than the Salpeter value (e.g., Harayama et al. 2008; Espinoza et al. 2009). Elmegreen \& Shadmehri (2003), Shadmehri (2004), Dib et al. (2007,2008), and Dib (2007) proposed that shallower-than Salpeter slopes can result from the efficient coalescence of closely packed protostellar cores in a dense protocluster environment (see also Huang et al. 2013).   

\section{CONCLUSIONS}\label{conclusions}

In this work, we test the universality of the IMF by comparing the fractions of single (i.e., isolated) and lonely O (single in their clusters and absence of massive B stars) stars in a sample of Galactic clusters and in synthetic cluster models constructed with various prior functions from which the parameters of the individual IMFs are randomly drawn. Using a Monte Carlo approach, the masses of stellar clusters are randomly sampled from an initial cluster mass function that is described by a power-law distribution. The IMF of stars within each cluster is randomly sampled using the tapered power-law (TPL) mass function. In order to make the synthetic clusters directly comparable to the observations, each cluster is assigned an age which is randomly drawn from an age distribution function similar to the observations, and are corrected for the effects of binary population and stellar evolution. Different models are constructed in which the set of three parameters that describe the TPL-IMF (i.e., the slope at the high mass end, $\Gamma$, the slope at the low mass end $\gamma$, and the characteristic mass, $M_{ch}$), assigned to each cluster are randomly sampled from parent distributions of varying widths, going from delta functions corresponding to the case of a universal IMF to broad distributions of the IMF parameters. After correcting for the effect of incompleteness, we compare the fractions of single and lonely O stars in these various models of simulated clusters with the fractions of single and lonely O stars measured for the population of young stellar clusters in the Milky Way.

Our work shows that the distributions of parameters that describe the IMF in a population of Milky Way stellar clusters are sufficiently broad such as to cast doubt on the idea of a universal and invariant IMF. Broad distributions of the parameters that describe the shape of the IMF are required in order to better reproduce the observed fractions of single and lonely O stars in the Milky Way stellar clusters. These broad distributions are compatible with the scatter between the IMF parameters of a more limited number of clusters found recently by Dib (2014a). We show that narrow distributions of the IMF parameters that are associated with the concept of a universal IMF are not favored by our results. Furthermore, our results suggest that star formation in clusters is stochastic and do not lend support to the existence of a deterministic cluster mass-maximum stellar mass relation. When the IMF is described by the tapered power-law, we propose that the parameters of the probabilistic IMF ($\Gamma$, $M_{ch}$, $\gamma$) be described by Gaussian probability distributions with the following standard deviations $\sigma_{\Gamma_{obs}}=0.6$, $\sigma_{M_{ch,obs}}=0.27$ M$_{\odot}$, $\sigma_{\gamma_{obs}}=0.25$, and centered around $\Gamma_{obs}=1.37$, $M_{ch,obs}=0.41$ M$_{\odot}$, and $\gamma_{obs}=0.91$, respectively. Future large and more sensitive Galactic and extragalactic surveys of stellar clusters will allow us to infer more accurately the shape of the distribution function of each of the IMF parameters.

The broad distributions of the IMF parameters inferred in this work very likely reflect the existence of equally broad distributions for the initial conditions under which these clusters have formed in Galactic proto-cluster clumps (e.g., Svoboda et al. 2016 ). As such, they offer an important motivation to explore physical mechanisms that can cause the IMF to vary from one star-forming region to another (e.g., different levels of mean accretion rates onto protostars, mergers of protostars, and the effects of feedback and triggering). The implications of our results are manifold. For example, the probabilistic IMF proposed in this work, in lieu of a invariant IMF, is expected to influence the modeling of star formation and stellar feedback in sub-grid models that are employed to describe star formation in local star forming regions in galactic and cosmological simulations. Broad distributions of the IMF parameters imply less mechanical and radiative feedback and chemical enrichment in local star forming regions with a steep slope of the IMF in the high mass regime versus more feedback and chemical enrichment in star forming regions with a shallow slope in this mass regime. 

\section*{Acknowledgments}

We thank the Referee for a thorough and careful reading of the manuscript and for constructive suggestions that helped improve and clarify several aspects of the paper. S. D. is supported by a Marie-Curie Intra European Fellowship under the European Community's Seventh Framework Program FP7/2007-2013 grant agreement no 627008. This work was supported by a research grant (VKR023406) from the Villum Foundation. S. S. was supported by Sonderforschungsbereich SFB 881 ÒThe Milky Way SystemÓ (subproject B5) of the German Research Foundation (DFG). S. H. acknowledges financial support from DFG programme HO 5475/2-1. This research has made use of NASA's Astrophysics Data System Bibliographic Services.
  
{}

\appendix 

\section{Sampling technique and additional details on the generated samples}\label{appendixa}

\begin{figure}
\begin{center}
\epsfig{figure=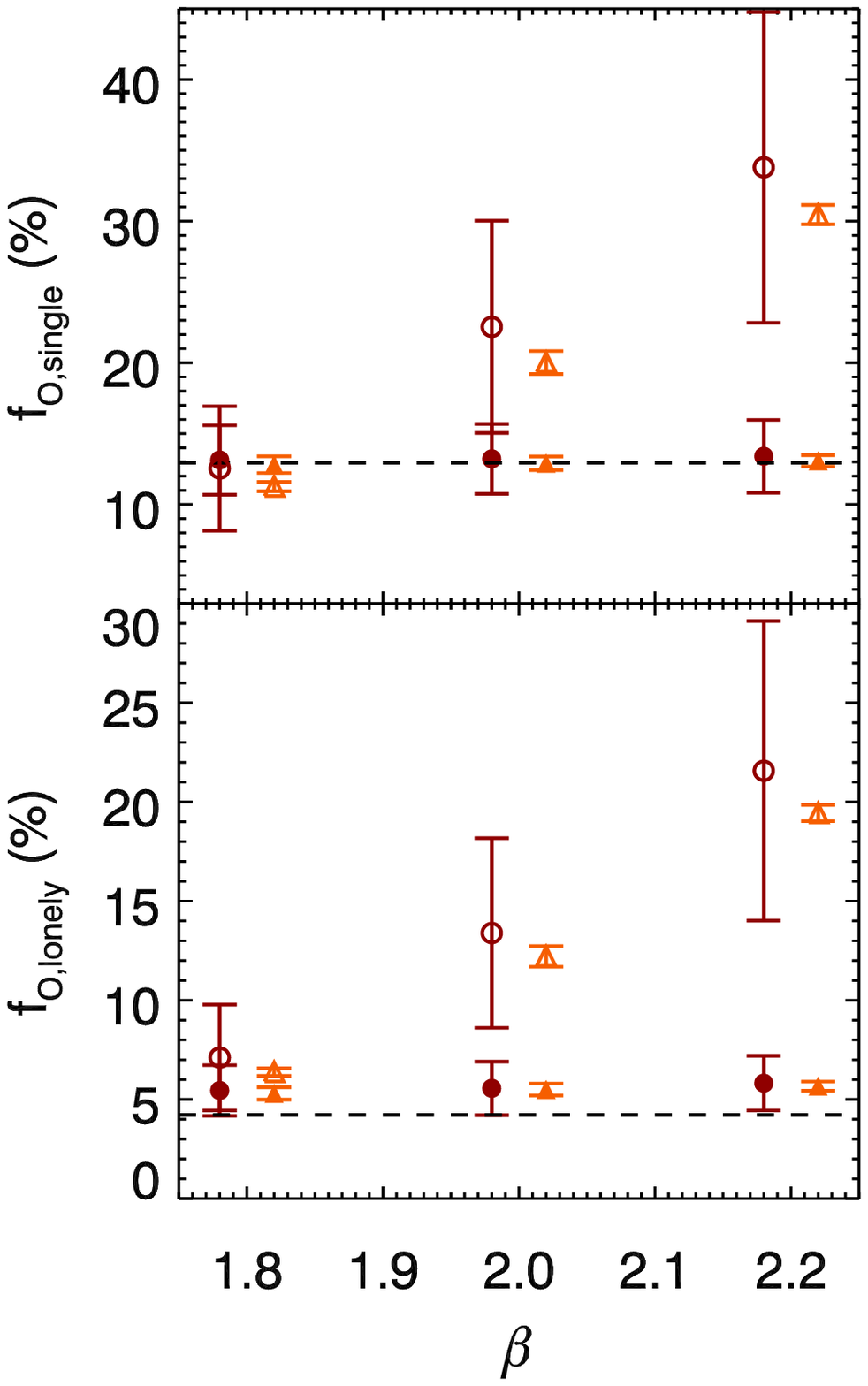,width=0.75\columnwidth}
\end{center}
\vspace{1.5cm}
\caption{Same as Fig.~\ref{fig8} (panel B) for the cases with completeness correction (filled symbols). The empty symbols show the same cases when no completeness correction is applied.}
\label{f1app}
\end{figure}

\begin{figure*}
\begin{center}
\epsfig{figure=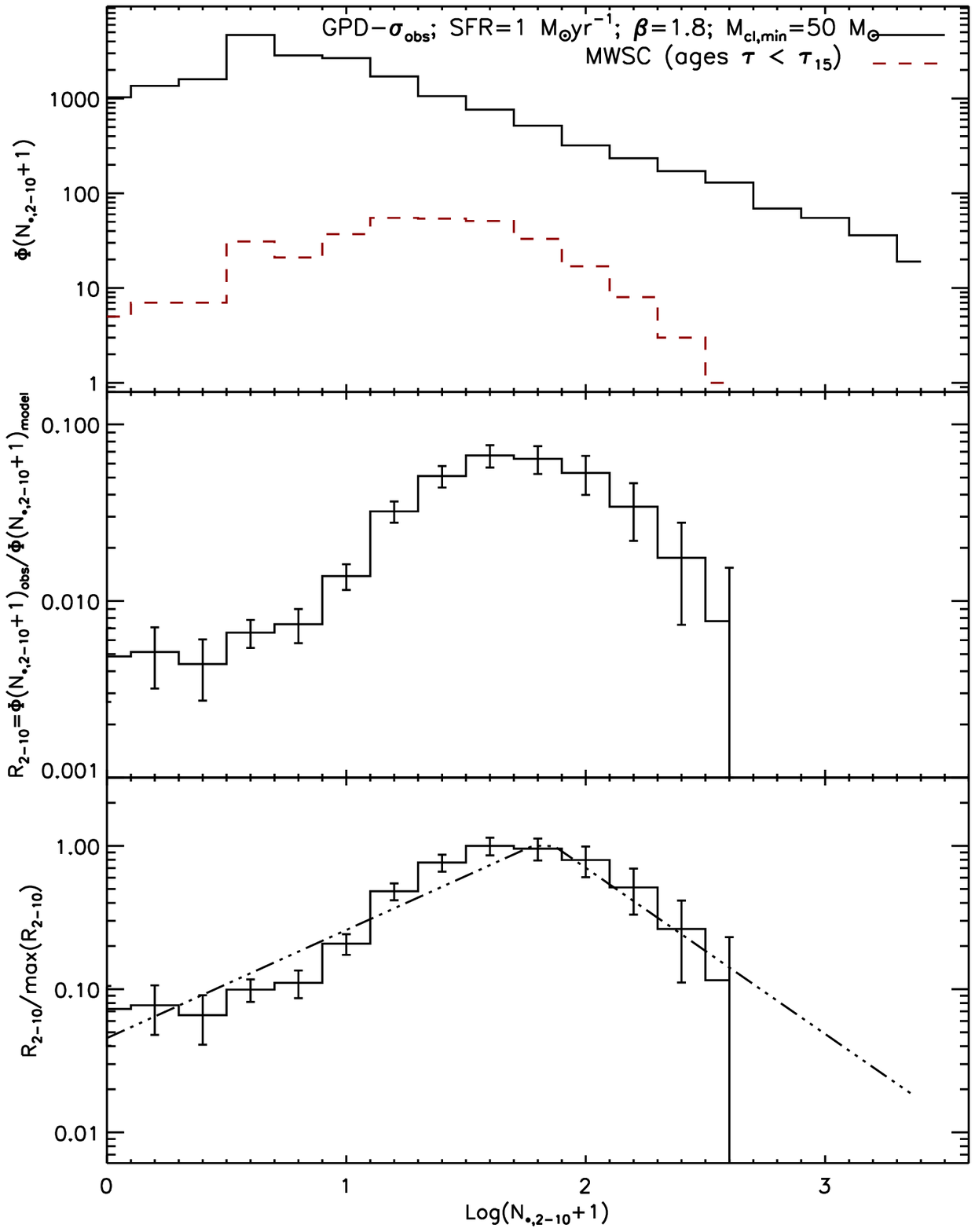,width=0.45\textwidth}
\hspace{0.5cm}
\epsfig{figure=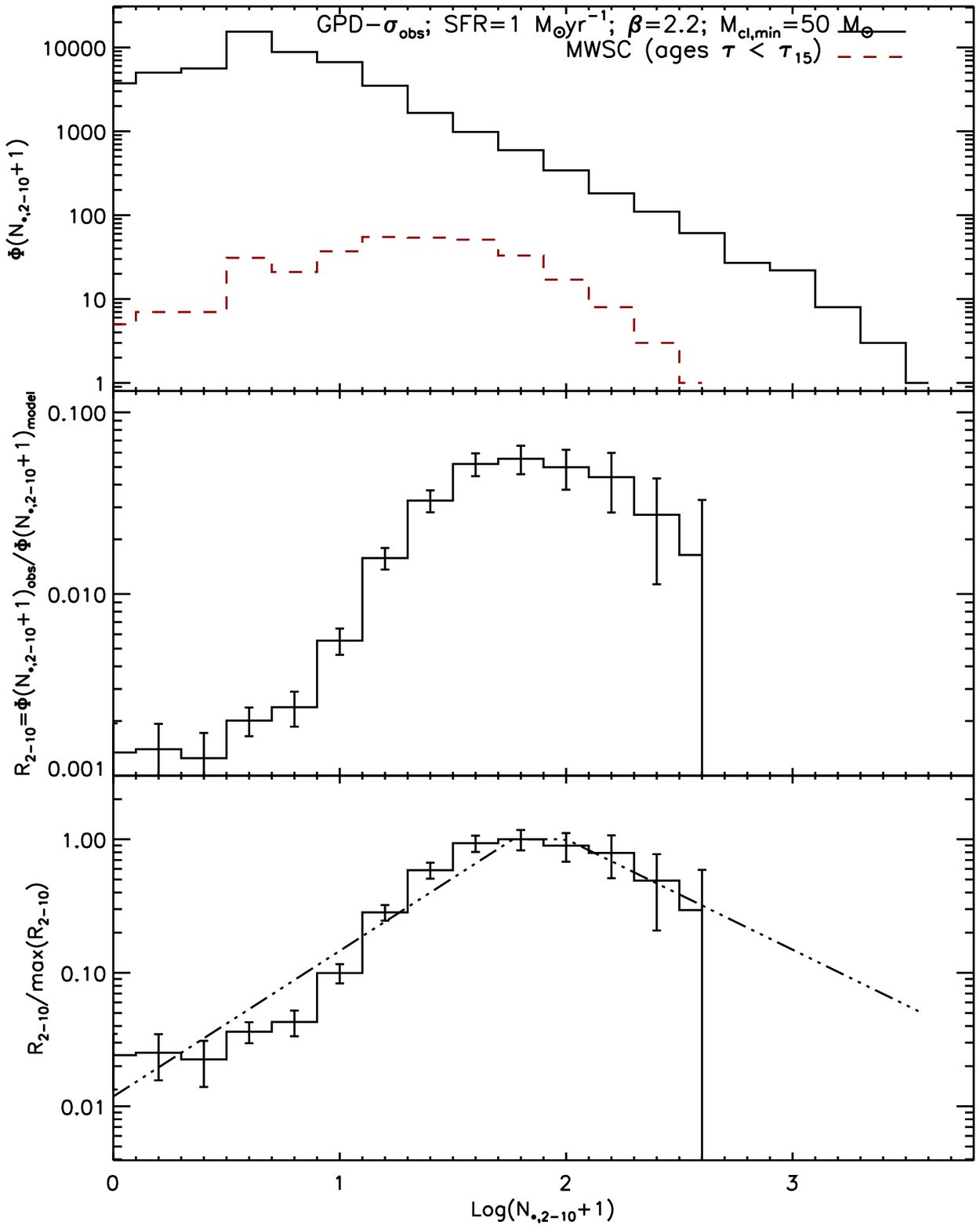,width=0.45\textwidth}
\end{center}
\vspace{0.5cm}
\caption{Same as Fig.~\ref{fig6} (left panel), but for a case with $\beta=1.8$ (left panel), and $\beta=2.2$ (right panel).}
\label{f2app}
\end{figure*}

The results presented in the main part of the paper rely on the random sampling of the masses of clusters in the initial cluster mass function (ICLMF) and on the masses of star systems in each individual cluster. Each cluster mass, $M_{cl}$, out of which stellar masses are randomly drawn, is itself drawn from a total mass reservoir given by $\Sigma_{cl}$=SFR$\times \tau_{15}$ where SFR is the assumed Galactic star formation rate and $\tau_{15}\approx12.3$ Myrs is sum of the Hydrogen+Helium burning phases of an O star with a mass of 15 M$_{\odot}$ (Ekstr\"{o}m et al. 2012). With the values of the Galactic SFR adopted in this work ($0.68$ to $1.45$ M$_{\odot}$ yr$^{-1}$), this yields mass reservoirs in the range $\approx \left(8.3-17\right) \times 10^{6}$ M$_{\odot}$. The methodology of the random sampling of cluster masses and of stellar masses is identical in nature with the exception that the probability distribution functions of the initial cluster mass function (ICLMF) is a power law function (Eq.~\ref{eq4}), whereas the system IMF is described by the tapered-power function (Eq.~\ref{eq6}) with a given set of parameters $\left(\Gamma,M_{ch}, \gamma\right)$ which are themselves randomly drawn from the different families of probability distribution functions displayed in Fig~\ref{fig1}.

In order to sample one mass from the ICLMF (respectively the IMF), we choose a uniform random value $M_{cl,i}$ (respectively, $M_{*,i}$) of the cluster mass (respectively, stellar mass) between the minimum and maximum mass limits assigned to each function ($M_{cl,min}$ and $M_{cl,max}$ for the ICLMF, and $M_{*,min}$ and $M_{*,max}$ for the IMF) as well as a random uniform value $Y_{i}$ using a standard random number generator between the minimum and maximum values of the ICLMF (respectively of the IMF) in the interval range of $M_{cl,min}$ and $M_{cl,max}$ for the ICLMF (and of $M_{*,min}$ and $M_{*,max}$ for the IMF). We evaluate ICLMF($M_{cl,i}$) (respectively, IMF($M_{*,i}$) and compare it to $Y_{i}$. If ICLMF($M_{cl,i}$) $> Y_{i}$ (respectively, if IMF($M_{*,i}$) $> Y_{i}$), the value of $M_{cl,i}$ (respectively, $M_{*,i}$) is admitted. Otherwise, the drawn mass is discarded and the sampling proceeds using a new value of $M_{cl,i}$ (respectively, of $M_{*,i}$). In theory, this iterative process should continue until the sum of the sampled masses is equal to $\Sigma_{cl}$ (or to $M_{cl}$ for the case of the IMF). However, the probability of the sum of sampled masses being exactly equal to the desired mass is marginal. It is therefore necessary to define a strategy for when to stop the sampling process. One method is the {\textquotedblleft stop after\textquotedblright} approach, in which the sampling of new masses is stopped immediately after the iteration that causes the sampled mass to be larger than the desired mass. An alternative is to remove the last mass that is drawn. In this {\textquotedblleft stop before\textquotedblright} approach, the total sampled mass is always smaller than the desired mass. As in Haas \& Anders (2010), we follow a {\textquotedblleft stop nearest\textquotedblright} approach in which we compare the sum of the sampled masses before and after the last iteration that causes the sum of sampled masses to go beyond the desired mass. Between these two iterations, the one that is adopted is the one that causes the sum of the sampled masses to be the nearest to the desired mass. 

The total number of clusters in the ICLMF depends on the adopted Galactic SFR, the chosen slope of the ICLMF, $\beta$, and on its lower mass cutoff $M_{cl,min}$ (the upper mass cutoff is fixed in our models to $5\times10^{4}$ M$_{\odot}$). With our adopted range of values for the SFR, $\beta$, and $M_{cl,min}$, the number of clusters in the ICLMF varies from $\approx 15000$ for the cases with the lowest value of the SFR ($0.68$ M$_{\odot}$ yr$^{-1}$), the lowest value of $\beta$ ($1.8$), and the largest value of $M_{cl,min}$ ($50$ M$_{\odot}$), to $\approx 80000$ for cases with the highest value of the SFR (1.45 M$_{\odot}$ yr$^{-1}$), the highest value of $\beta$ ($2.2$), and for $M_{cl,min}=50$ M$_{\odot}$, and up to $\approx 210000$ for the highest values of the SFR (1.45 M$_{\odot}$ yr$^{-1}$), the highest value of $\beta$ (2.2), and the lowest value of $M_{cl,min}$ ($10$ M$_{\odot}$). It should be noted that since the quantities we are calculating ($f_{O,single}$ and $f_{O,lonely}$) are dimensionless numbers, the calculated values are insensitive to the choice of the SFR, insofar as the ICLMF is complete at the low mass end, which is the case for the values of parameters explored in this work. 
 
\section{Additional examples of the completeness function} \label{appendixb}

In \S.~\ref{complete}, we described the method for constructing the completeness correction function for each realization of the ICLMF. Fig.~\ref{fig6} (left panel) displays an example of the completeness function for a case with the GPD-$\sigma_{obs}$ distribution of the parameter and with the following set of other parameters, ${\rm SFR}=1$ M$_{\odot}$ yr$^{-1}$, $M_{cl,min}=50$ M$_{\odot}$, and the slope of the ICLMF, $\beta=2$. Here, we discuss how the shape of the completeness function is affected by the shape of the ICLMF and how this impacts the derived values of $f_{O,single}$ and $f_{O,lonely}$ with respect to the case where no completeness correction is applied (i.e., case of full completeness). As discussed in \S.~\ref{complete}, in the absence of any completeness correction, the values of $f_{O,single}$ and $f_{O,lonely}$ increase with increasing values of $\beta$. A steeper ICLMF implies a larger fraction of low mass clusters which are more likely to harbor single and lonely O stars. In contrast, a shallow ICLMF is more deficient in low mass clusters, and this leads to a reduction of the value of $f_{O,single}$ and $f_{O,lonely}$. This is illustrated in Fig.~\ref{f1app} which displays the values of $f_{O,single}$ and $f_{O,lonely}$ as a function of $\beta$ without any completeness correction (open symbols) and of the case where the completeness correction is applied (full symbols). 

This trend of increasing values of $f_{O,single}$ and $f_{O,lonely}$ with increasing values of $\beta$ are washed away by the application of the completeness corrections. The reason for this effect lies in the fact that a steep ICLMF ($\beta=2.2$) results in a shallower $f_{comp}$ function in comparison with the case with $\beta=2$ at high values of $N_{*,2-10}$ (i.e., for high mass clusters) and to a steeper $f_{comp}$ function at low values of $N_{*,2-10}$ (i.e., for low mass clusters). An example of the completeness function for a realization of the ICLMF with $\beta=2.2$ is displayed in Fig.~\ref{f2app} (right panel). In comparison to cases with lower values of $\beta$, such a distribution of $f_{comp}$ for a steep ICLMF leads to a larger relative retention of massive clusters that harbor large numbers of massive stars and this in turn leads to a significant reduction of the value of $f_{O,single}$ and $f_{O,lonely}$ when compared to the case with no completeness correction. In contrast, smaller values of $\beta$ result in a steeper $f_{comp}$ function at high values of $N_{*,2-10}$ and to a shallower $f_{comp}$ function at low values of $N_{*,2-10}$ (case for $\beta=1.8$ in Fig.~\ref{f2app}, left panel). This leads to the dismissal of a relatively larger fraction of massive clusters and while the net effect is still a reduction of $f_{O,single}$ and $f_{O,lonely}$ with respect to case with full completeness, the effect becomes less pronounced at lower values of $\beta$.

\label{lastpage}


\begin{thebibliography}{}

\bibitem[Alvesdeoliveira (2013)] {alvesdeoliveira13} Alves de Oliveira, C., Moraux, E., Bouvier, J., et al. 2013, A\&A, 549, 123  
\bibitem[Ascenco (2007)] {ascenco07} Ascenco, J., Alves, J., Vicente, S., Lago, M. T. V. T. 2007, A\&A, 476, 199
\bibitem[Basu (2015)] {basu15} Basu, S., Gil, M., Auddy, S. 2015, MNRAS, 449, 2413
\bibitem[Bochanski (2010)] {bochanski10} Bochanski, J. J., Hawley, S. L., Covey, K. R., West, A. A., Reid, I. N., Golimowski, D. A., Ivezi\'{c}, Z. 2010, AJ, 139, 2679
\bibitem[Boissier (1999)] {boissier99} Boissier, S., Prantzos, N. 1999, MNRAS, 307, 857
\bibitem[Bouvier (2008)] {bouvier08} Bouvier J., Kendall, T., Meeus, G. et al. 2008, A\&A, 481, 661
\bibitem[Carpenter (2000)] {carpenter00} Carpenter, J. M. 2000, MNRAS, 120, 3139
\bibitem[Cervino (2006)] {cervino06} Cervi\~{n}o, M., Luridiana, V. 2006, A\&A, 451, 475
\bibitem[Cervino (2009)] {cervino09} Cervi\~{n}o, M., Luridiana, V. 2009, in New Quests in Stellar Astrophysics. II. Ultraviolet Properties of Evolved Stellar Populations, ed. M. Ch\'{a}vez Dagostino, E. Bertone, D. Rosa Gonzalez, \& L. H. Rodriguez-Merino (Dordrecht: Springer), 293
\bibitem[Cervino (2013)] {cervino13} Cervi\~{n}o, M. 2013, New Astron. Rev., 57, 123 
\bibitem[Chabrier (2003)] {chabrier03} Chabrier, G. 2003, PASP, 115, 763
\bibitem[Chabrier  (2005)] {chabrier05} Chabrier, G. 2005, in Astrophysics and Space Science Library, Vol. 327, The Initial Mass Function 50 Years Later, ed. E. Corbelli, F. Palla, \& H. Zinnecker, 41
\bibitem[Chandar (2012)] {chandar12} Chandar, R., Fall, S. M., Whitmore, B. C. 2010, ApJ, 711, 1263
\bibitem[Chini (2012)] {chini12} Chini, R., Hoffmeister, V. H., Nasseri, A., Stahl, O., Zinnecker, H. 2012, MNRAS, 424, 1925  
\bibitem[Clark (2009)] {clark09} Clark, J. S., Negueruela, I., Davies, B., Larionov, V. M., Richie, B. W. et al. 2009, A\&A, 498, 109
\bibitem[Conroy (2009)] {conroy09} Conroy, C., Gunn, J. E., White, M. 2009, ApJ, 699, 486
\bibitem[da Silva (2012)] {dasilva12} da Silva, R. L., Fumagalli, M., Krumholz, M. R. 2012, ApJ, 745, 145
\bibitem[da Silva (2014)] {dasilva14} da Silva, R. L., Fumagalli, M., Krumholz, M. R. 2014, MNRAS, 444, 3275
\bibitem[de Grijs (2006)] {degrijs06} de Grijs, R., Anders, P. 2006, MNRAS, 366, 295 
\bibitem[Delgado (2011)] {delgado11} Delgado, A. J., Alfaro, E., J., Yun, J. L. 2011, A\&A, 531, 141 
\bibitem[De Marchi (2010)] {demarchi10} De Marchi, G., Paresce, F., Portegies Zwart, S. 2010, ApJ, 718, 23
\bibitem[de Wit (2005)] {dewit05} de Wit, W. J., Testi, L., Palla, F., Zinnecker, H. 2005, A\&A, 437, 247
\bibitem[Dib (2006)] {dib06} Dib, S., Bell, E., Burkert, A. 2006, ApJ, 638, 797
\bibitem[Dib (2007)] {dib07a} Dib, S. 2007, JKAS, 40, 157
\bibitem[Dib (2007)] {dib07b} Dib, S., Kim, J., Shadmehri, M. 2007, MNRAS, 381, L40 
\bibitem[Dib (2008)] {dib08} Dib, S., Shadmehri, M., Gopinathan, M., Kim, J., Henning, Th. 2008, in Massive Star Formation: Observations confront Theory, ASP Conf. Series, Ed. H. Beuther, H. Linz, T. Henning, 387, 282
\bibitem[Dib (2010)] {dib10} Dib, S., Shadmehri, M., Padoan, P., Maheswar, G., Ojha, D. K., Khajenabi, F. 2010, MNRAS, 405, 401
\bibitem[Dib (2011a)] {dib11a} Dib, S., Piau, L., Mohanty, S., Braine, J. 2011a, MNRAS, 415, 3439
\bibitem[Dib (2011b)] {dib11b} Dib, S., Piau, L., Mohanty, S., Braine, J. 2011b, in SF2A-2011, Proceedings of the Annual meeting of the French Society of Astronomy and Astrophysics, ed. G, Alicia, K. Belkacem, R. Samadi, \& D. Valls-Gabaud (Paris: Societe Francaise d'Astronomie et d'Astrophysique), 275
\bibitem[Dib (2011)] {dib11} Dib, S. 2011, in Stellar Clusters \& Associations: A RIA Workshop on GAIA, Eds. E. J. Alvaro Navarro, A. T. Gallego Calvente, M. R. Zapatero Osorio, 30
\bibitem[Dib (2013)] {dib13} Dib, S., Gutkin, J., Brandner, W., Basu, S. 2013, MNRAS, 436, 3727
\bibitem[Dib (2014)] {dib14a} Dib, S. 2014a, MNRAS, 444, 1957
\bibitem[Dib (2014)] {dib14b} Dib, S. 2014b, in The Labyrinth of Star Formation, ed. D. Stamatellos, S. Goodwin, \& D. Ward-Thompson (Springer)
\bibitem[Ekstrom (2012)] {ekstrom12} Ekstr\"{o}m, S., Georgy, C., Eggenberger, P., Meynet, G., Mowlavi, N. et al. 2012, A\&A, 537, 136
\bibitem[Eldridge (2009)] {eldridge09} Eldridge, J. J., Stanway, E. R. 2009, MNRAS, 400, 1019
\bibitem[Elmegreen (1997)] {elmegreen97} Elmegreen, B. G., Efremov, Y. N. 1997, ApJ, 480, 235
\bibitem[Elmegreen (2003)] {elmegreen03} Elmegreen, B. G., Shadmehri, M. 2003, MNRAS, 338, 817
\bibitem[Elmegreen (2004)] {elmegreen04} Elmegreen, B. G. 2004, MNRAS, 354, 367
\bibitem[Elmegreen (2008)] {elmegreen08} Elmegreen, B. G., Klessen, R. S., Wilson, C. D. 2008, ApJ, 681, 365
\bibitem[Espinoza (2009)] {espinoza09} Espinoza, P., Selman, F. J., Melnick, J. 2009, A\&A, 501, 563
\bibitem[Fall (2012)] {fall12} Fall, S. M., Chandar, R. 2012, ApJ, 752, 96    
\bibitem[Figer (1999)] {figer99} Figer, D. F., McLean, I. S., Morris, M. 1999, ApJ, 514, 202
\bibitem[Fumagalli (2011)] {fumaggali11} Fumagalli, D., da Silva, R. K., Krumholz, M. R. 2011, ApJ, 741, 26
\bibitem[Gennaro (2012)] {gennaro12} Gennaro, M., Brandner, W., Stolte, A., Henning, Th. 2011, MNRAS, 412, 2469
\bibitem[Girardi (2000)] {girardi00} Girardi, L., Bertelli, A., Bressan, C., Chiosi, C., Groenewegen, M. A. T. et al. 2002, A\&A, 391, 195   
\bibitem[Haas (2010)] {haas10} Haas, M. R., Anders, P. 2010, A\&A, 512, 79   
\bibitem[Harayama (2008)] {harayama08} Harayama, Y., Eisenhauer, F., Martins, F. 2008, ApJ, 675, 1319
\bibitem[Hermanowicz (2011)] {hermanowicz11} Hermanowicz, M. T., Kennicutt, R. C., Aldridge, J. J.  2013, MNRAS, 432, 3097  
\bibitem[Hony (2015)] {hony15} Hony, S., Gouliermis, D. A., Galliano, F., Galametz, M., Cormier, D., Chen, C.-H. R., Dib, S. et al. 2015, MNRAS, 448, 184
\bibitem[Huang (2013)] {huang13} Huang, X., Zhou, T., Lin, D. N. C. 2013, ApJ, 769, 23
\bibitem[Hunter (2005)] {hunter05} Hunter, D. A., Elmegreen, B. G., Dupuy, T. J., Mortonson, M. 2003, AJ, 126, 1836 
\bibitem[Kharchenko (2012)] {kharchenko12} Kharchenko, N. V., Piskunov, A. E., Schilbach, E., R\"{o}ser, S., Scholz, R.-D. 2012, A\&A, 543, 156
\bibitem[Kharchenko (2013)] {kharchenko13} Kharchenko, N. V., Piskunov, A. E., Schilbach, E., R\"{o}ser, S., Scholz, R.-D. 2013, A\&A, 558, 53
\bibitem[Kobulnicky (2015)] {kobulnicky15} Kobulnicky, H. A., Kimiki, D. C., Lundquist, M. J., Burke, J., Chapman et al. 2014, ApJS, 213, 34
\bibitem[Kroupa (1993)] {kroupa93} Kroupa, P., Tout, C. A., Gilmore, G. 1993, MNRAS, 262, 545
\bibitem[Kroupa (2001)] {kroupa01} Kroupa, P. 2001, MNRAS, 322, 231  
\bibitem[Krumholz (2016)] {krumholz16} Krumholz, M. R., Myers, A. T., Klein, R. I., McKee, C. F. 2016, MNRAS, 460, 3272
\bibitem[Lada (2003)] {lada03} Lada, C. J., Lada, E. A. 2003, ARA\&A , 41, 57
\bibitem[Lamb (2010)] {lamb10} Lamb, J. B., Oey, M. S., Werk, J. K., Ingleby, L. D. 2010, ApJ, 725, 1886 
\bibitem[Larsen (2009)] {larsen09} Larsen, S. S. 2009, A\&A, 494, 539 
\bibitem[Lim (2005)] {lim05} Lim, B., Sung, H., Hur, H., Park, B.-G., 2015, in the proceedings of IAU 316, Eds. C. Charbonnel, A. Nota, arXiv: 1511.01118
\bibitem[Liu (2009)] {liu09} Liu, Q., de Grijs, R., Deng, L. C., Hu, Y., Baraffe, I., Beaulieu, S. F. 2009, MNRAS, 396, 1665 
\bibitem[Liu (2003)] {liu03} Liu, W. M., Meyer, M. R., Cotera, A. S., Young, E. T. 2003, AJ, 126, 1665
\bibitem[Lodieu (2011)] {lodieu11} Lodieu, N., Dobbie, P. D., Hambly, N. C. 2011, A\&A, 527, 24 
\bibitem[Luhman (2004)] {luhman04} Luhman, K. L. 2004, ApJ, 617, 1216
\bibitem[Luhman (2007)] {luhman07} Luhman, K. L. 2007, ApJ, 173, 104 
\bibitem[Maia (2016)] {maia16} Maia, F. F. S., Moraux, E., Joncour, I. 2016, MNRAS, 458, 3027
\bibitem[Mallick (2014)] {mallick14} Mallick, K. K., Ojha, D. K., Tamura, M., Pandey, A. K., Dib, S. et al. 2014, MNRAS, 443, 3218
\bibitem[Marigo (2008)] {marigo08} Marigo, P., Girardi, L., Bressan, A., et al. 2008, A\&A, 482, 883
\bibitem[Martizzi (2016)] {martizzi16} Martizzi, D., Fielding, D., Faucher-Gigu\`{e}re, C.-A., Quataert, E. 2016, MNRAS, 459, 2311
\bibitem[Maschberger (2008)] {maschberger08} Maschberger, T., Clarke, C. J. 2008, MNRAS, 391, 711
\bibitem[Maschberger (2013)] {maschberger13} Maschberger, T. 2013, MNRAS, 429, 1725 
\bibitem[Massey (1998)] {massey98a} Massey, P. 1998, in the Stellar Initial Mass Function, Eds, G. Gilmore, \& D. Howell (San Francisco: ASP), vol 142 of ASP Conf. Ser. 17
\bibitem[Massey (2011)] {massey10} Massey, P. 2011, in the proceedings of UP2010: Have Observations Revealed a Variable Upper End of the Initial Mass Function?  M. Treyer, T.K. Wyder, J.D. Neill et al. (eds.) (San Francisco: ASP), ASP Conf. Ser 440, p. 29
\bibitem[Miller (1979)] {miller79} Miller, G. E., Scalo, J. M. 1979, ApJS, 41, 513
\bibitem[Moraux (2004)] {moraux04} Moraux, E., Kroupa, P., Bouvier, J. 2004, A\&A, 426, 75
\bibitem[Oh (2015)] {oh15} Oh, S., Kroupa, P., Pflamm-altenburg, J. 2015, ApJ, 805, 92
\bibitem[Ojha (2010)] {ojha10} Ojha, D. K., Kumar, M. S. N., Davis, C. J., Grave, J. M. C. 2010, MNRAS, 407, 1807
\bibitem[Parker (2007)] {parker07} Parker, R. J., Goodwin, S. P. 2007, MNRAS, 380, 1271
\bibitem[Parravano (2011)] {parravano11} Parravano, A., McKee, C. F., Hollenbach, D. J. 2011, ApJ, 726, 27
\bibitem[Preibisch (2002)] {preibisch02} Preibisch, T., Brown, A. G. A., Bridges, T., Gunether, E., Zinnecker, H. 2002, AJ, 124, 404
\bibitem[Robitaille (2010)] {robitaille10} Robitaille, T. P., Whitney, B. A. 2010, ApJ, 710, 11
\bibitem[Rybizki (2015)] {rybizki15} Rybizki, J., Just, A. 2015, MNRAS, 447, 3880
\bibitem[Salpeter (1955)] {salpeter55} Salpeter, E. E. 1955, ApJ, 121, 161 
\bibitem[Scalo (1986)] {scalo86} Scalo, J. M. 1986, Fundamentals of Cosmic Physics, 11, 1 
\bibitem[Scalo (1998)] {scalo98} Scalo, J. 1998, in The Stellar Initial Mass Function, ed. G. Gilmore, I. Parry, S. Ryan, Cambridge: Cambridge University Press, p. 201 
\bibitem[Scalo (2005)] {scalo05} Scalo, J. 2005, in The Initial Mass Function 50 years later. Eds by E. Corbelli and F. Palla, ASSL, 327, 23  
\bibitem[Schilbach (2006)] {schilbach06} Schilbach, E., Kharchenko, N. V., Piskunov, A. E., R\"{o}ser, S., Scholz, R.-D. 2006, A\&A, 456, 523
\bibitem[Schmeja (2014)] {schmeja14} Schmeja, S., Kharchenko, N. V., Piskunov, A. E., R\"{o}ser, S., Schilbach, E. 2014, A\&A,  568, 51
\bibitem[Scholz (2013)] {scholz13} Scholz, A., Geers, V., Clark, P., Jayawardhana, R., Muzic, K. 2013, ApJ, 775, 138
\bibitem[Selman (2005)] {selman05} Selman, F. J., Melnick, J. 2005, A\&A, 443, 851
\bibitem[Selman (2008)] {selman08} Selman, F. J., Melnick, J. 2008, ApJ, 689, 816
\bibitem[Shadmehri (2004)] {shadmehri04} Shadmehri, M. 2004, MNRAS, 354, 375
\bibitem[Sharma (2008)] {sharma08} Sharma, S., Pandey, A. K., Ogura, K. et al. 2008, AJ, 135, 1934    
\bibitem[Shatsky (2002)] {shatsky02} Shatsky, N., Tokovinin, A. 2002, A\&A, 382, 92
\bibitem[Sung (2010)] {sung10} Sung, H., Bessell, M. S. 2010, AJ, 140, 2070 
\bibitem[Vanbeveren (1982)] {vanbeveren82} Vanbeveren, D. 1982, A\&A, 115, 65
\bibitem[Vanbeveren (2009)] {vanbeveren09} Vanbeveren, D. 2009, New Astron. Rev., 53, 27
\bibitem[Vincke (2015)] {vincke15} Vincke, K., Breslau, A., Pfalzner, S. 2015, A\&A, 577, 115 
\bibitem[Weidner (2004)] {weidner04} Weidner, C., Kroupa, P. 2004, ApJ, 348, 187 
\bibitem[Weidner (2013)] {weidner13} Weidner, C., Kroupa, P., Pflamm-Altenburg, J. 2013, MNRAS, 434, 84
\bibitem[Weisz (2012)] {weisz12} Weisz, D. R., Johnson, B. D., Johnson, L. C., Skillman, E. D., Lee, R. C. et al. 2012, ApJ, 744, 44
\bibitem[Weisz (2015)] {weisz15} Weisz, D. R., Johnson, C. L., Foreman-Mackey, D.,  Dolphin, A. E., Beerman, B. F. et al. 2015, ApJ, 806, 198
\bibitem[Zhang (1999)] {zhang99} Zhang, Q., Fall, S. M. 1999, ApJ, 527, L81
\bibitem[Zinnecker (2007)] {zinnecker07} Zinnecker, H., Yorke, H. W. 2007, ARA\&A, 45, 481
 
\end{thebibliography}
\end{document}